\documentclass[universe,pdftex,moreauthors]{mdpi} 
\firstpage{1} 
\pubvolume{1}
\issuenum{1}
\usepackage{epstopdf}
\usepackage{threeparttable} 
\articlenumber{0}
\pubyear{2025}
\copyrightyear{2025}
\datereceived{2025} 
\daterevised{ } 
\dateaccepted{ } 
\datepublished{ } 
\hreflink{https://doi.org/} 
\Title{Millihertz quasi-periodic oscillations in accreting X-ray pulsars}
\TitleCitation{Millihertz quasi-periodic oscillations in accreting X-ray  pulsars}
\Title{Millihertz quasi-periodic oscillations in accreting X-ray pulsars}
\TitleCitation{Title}

\Author{Wen Yang $^{1,}$\orcidA{}, Wei Wang $^{1,*}$\orcidB{}}
\AuthorNames{Wen Yang and Wei Wang}
\isAPAStyle{%
       \AuthorCitation{Yang, W., \& Wang, W.}
         }{%
        \isChicagoStyle{%
        \AuthorCitation{Wen Yang and Wei Wang}
        }{
        \AuthorCitation{Yang, W.; Wang, W.}
        }
}
\address{%
$^{1}$ \quad Department of Astronomy, School of Physics and Technology, Wuhan University, Wuhan 430072, China}
\corres{Correspondence: wangwei2017@whu.edu.cn (Wei Wang)}
\abstract{Accreting neutron stars exhibit pulsed X-rays and complex temporal variability across multi-wavelengths and different timescales. This variability could be driven by various physical processes including instability or inhomogeneous motions within the accretion flow, thermonuclear bursts on the neutron star surface. In this review, we present a concise overview of the observational features for millihertz (mHz) quasi-periodic oscillations (QPOs) at a frequency range of $\sim 1- 1000$ mHz observed in light curves of X-ray pulsars for both low-mass X-ray binaries and high-mass X-ray binaries, based on recent X-ray missions, e.g., NICER, Insight-HXMT and NuSTAR. We further summarize current theoretical interpretations, discuss remaining challenges and propose potential directions for future studies to advance the understanding of the nature and physical origin of these QPOs.}
\let\linenumbers\relax
\begin{document}
\section{Introduction}
X-ray binary systems consist of a compact object (such as a neutron star (NS) or black hole (BH)) and an optical companion, with intense X-ray emission produced as the compact object accretes matter from the donor star \citep{reig2011x}. X-ray binaries can be classified into three categories based on the mass of the companion star ($M_{co}$) \citep{bradt1983optical}: low-mass X-ray binaries (LMXBs) with $M_{\mathrm{co}}<1 M_{\odot}$, intermediate-mass X-ray binaries (IMXBs) with $1 M_{\odot}<M_{\mathrm{co}}<8 M_{\odot}$ and high-mass X-ray binaries (HMXBs) with $M_{\mathrm{co}}>8 M_{\odot}$.
\par
\begin{figure}[h]
    \centering
    \includegraphics[width=0.88\textwidth]{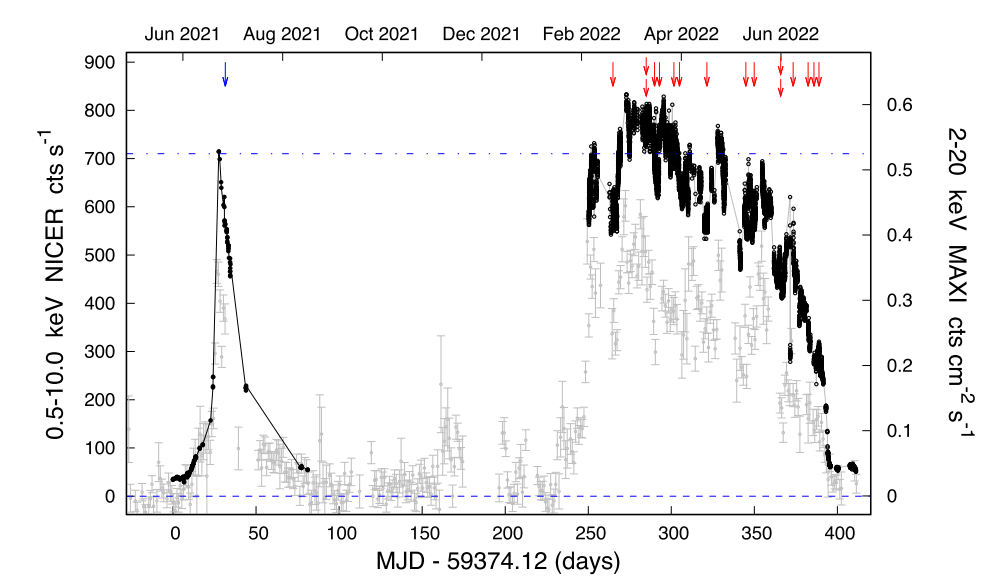}
    \caption{The temporal evolution of the 2021 and 2022 outbursts of 4U 1730–22. The blue arrow marks the single Type-I burst in 2021, while the red arrows mark all Type-I bursts detected in 2022. The vertical axis indicates the X-ray count rate. Data points from NICER (black data points, 0.5–10 keV) and with MAXI (grey data points, 2–20 keV) are shown. The figure is taken from \citep{mancuso2023detection}}. 
    \label{fig:fig1}
\end{figure}

In NS-LMXBs, the spectral type of the optical star is later than A \citep{van1997optical}. The spatial distribution concentrates toward the Galactic Center, in particular with high density towards the Galactic Bulge, where older stars are found \citep{sazonov2020galactic}. They are characterized by relatively weak surface magnetic fields of $10^8$–$10^9$ G and short spin periods of a few milliseconds, while their orbital periods span from minutes to approximately 20 days \citep{asai2022decades}. NS-LMXBs are commonly classified into Z sources and Atoll sources distinguished by their characteristic Z-shaped and atoll-shaped tracks in X-ray color–color diagrams (CCDs) or hardness–intensity diagrams (HIDs). Z sources exhibit high luminosities near the Eddington luminosity ($L_E$) and trace out a Z-shaped track in the CCD, which is composed of the horizontal branch (HB), normal branch (NB), and flaring branch (FB) \citep{hasinger1989two}. Atoll sources span a much broader luminosity range, from 0.001 $L_E$ up to the characteristic luminosity levels of Z sources. In CCDs, they typically exhibit a C-shaped or U-shaped track, consisting of two primary branches: the island branch, which often appears as discrete patches, and the banana branch, which forms a more continuous structure \citep{schnerr2003peculiar}. Type I X-ray bursts, characterized by spectral signatures of photospheric cooling, have been observed in LMXBs for more than three decades \citep{grindlay1976discovery,belian1976discovery} and recognized as thermonuclear bursts occurring in accreting NS \citep{lewin1977x,woosley1976gamma}. Figure \ref{fig:fig1} shows the temporal evolution of the 2021 and 2022 outbursts of 4U 1730-22, which included multiple Type I X-ray bursts.
\par
\begin{figure}[h]
    \centering
    \includegraphics[width=0.45\textwidth]{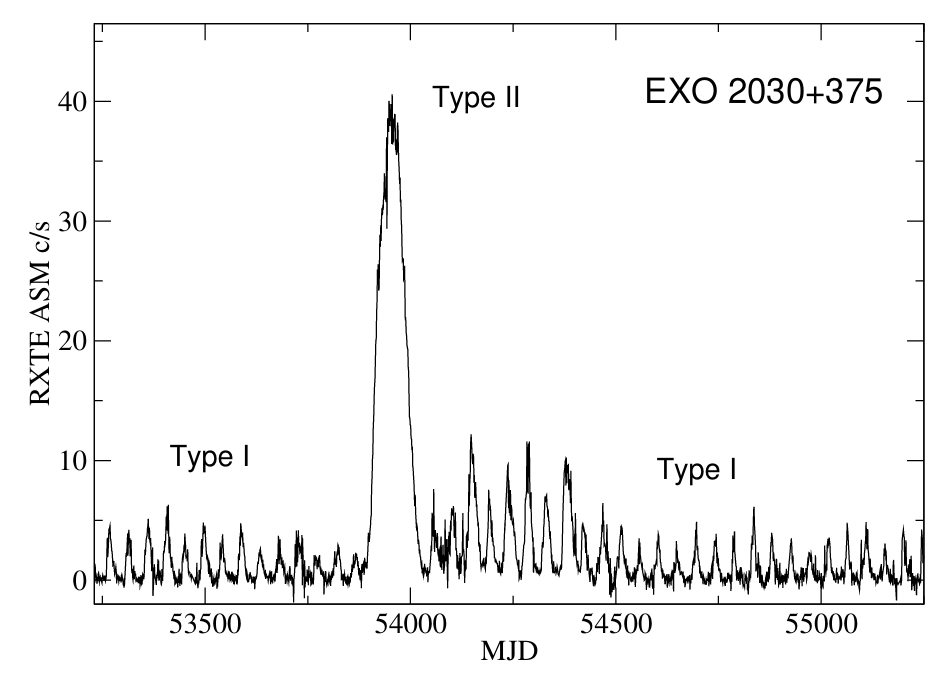}
    \includegraphics[width=0.47\textwidth]{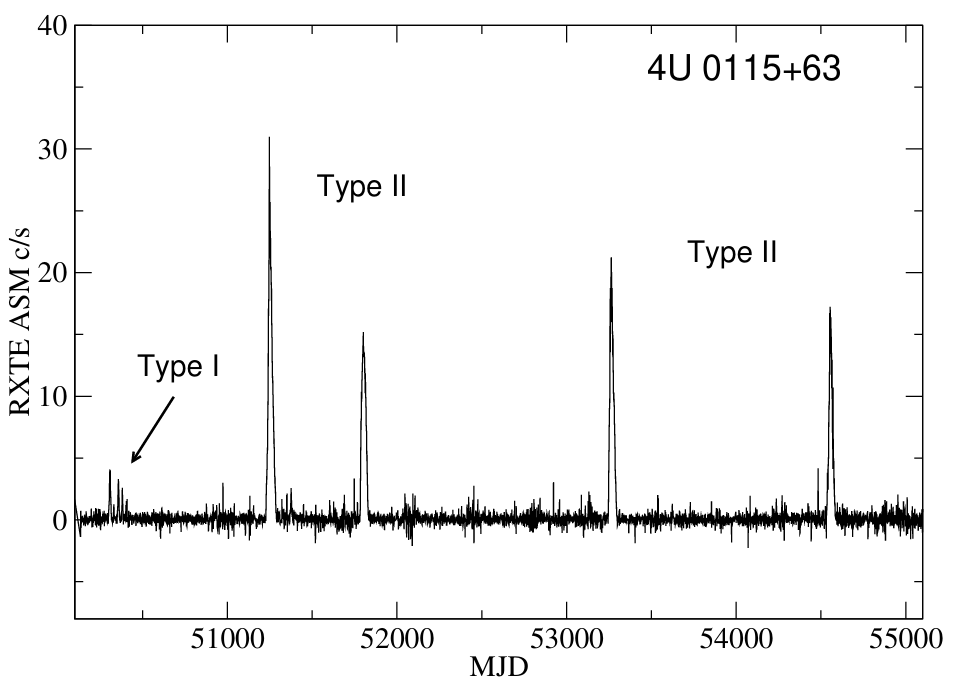}
    \caption{Long-term X-ray light curves of EXO 2030+375 (left) and 4U 0115+63 (right), taken from \citep{reig2011x}. The vertical axis indicates the X-ray count rate. EXO 2030+375 shows one rare Type II outburst and frequent regular Type I outbursts, while 4U 0115+63 exhibited several Type I outbursts, and four Type II outbursts between 1996 and 2009. }
    \label{fig:fig2}
\end{figure}

NS-HMXBs contain early-type (O or B) companions and their spatial density is distributed along the Galactic plane, reflecting the star formation rate of the host galaxy \citep{liu2006catalogue,lutovinov2013population}. HMXBs with strong magnetic field neutron stars are among the most intense sources of X-ray emission in the sky. Their high-energy radiation originates from the accretion process, in which the flow of infalling gas is channeled along magnetic field lines to the surface of a neutron star which is less than the whole surface area of the star, creating localized “hot spots” where its kinetic energy is transformed into radiation \citep{davidson1973accretion}. Since the magnetic axis may be misaligned with the rotation axis, as the star rotates, the radiation from the hot spots forms a pulse profile with the rotation period in the range 1-1000 s \citep{dai2006exploration,davies1979spindown,zhang2004spin}. Cyclotron resonant scattering features (CRSFs) at the energies of $\sim 10 -100$ keV are discovered in hard X-ray spectra of these sources which confirm the strongly magnetized neutron stars ($10^{12}-10^{13} $ G) located in accreting binary systems \citep{mukherjee2014magnetic,liuq2022vela,yangw2023cen,yangw2024exo2030}. The luminosity features are used to categorize NS-HMXBs into two groups: Be/X-ray binaries (BeXBs), which contain optical stars of luminosity class III–V (dwarfs, subgiants, or giant OBe stars), and SGXBs, which contain luminosity class I–II optical stars \citep{reig2011x}. Most BeXBs are transient sources with orbital eccentricities typically exceeding $\sim 0.3$. In these systems, X-ray emission arises when the compact object accretes matter from a quasi-Keplerian equatorial disk surrounding the rapidly rotating Be star. Such a mechanism explains normal (Type I) outbursts with X-ray luminosity $L_x\,\,\sim~10^{35}-10^{37}\,\,erg\,\,s^{-1}$. Type II X-ray bursts are major events that represent a peak luminosity higher than $10^{38}{\,\mathrm{erg~s^{-1}}}$ and are normally known to last for several weeks to months \citep{okazaki2001natural}. Figure \ref{fig:fig2} shows these two types of X-ray variability of BeXBs for EXO 2030+375 and 4U 0115+63. In classical SGXBs, powerful and continuous stellar winds from the supergiant companion, with mass-loss rates of $10^{-6}$–$10^{-8}M_{\odot}{\rm \ yr^{-1}}$, provide a steady supply of material accreted onto the compact object, leading to relatively stable X-ray emission with typical luminosities of $L_X \sim 10^{35}$–$10^{37}{\rm erg \ s^{-1}}$ \citep{di2025fast}. 
\par
Over the past four decades, QPOs as prominent timing features have been widely identified in accreting X-ray binaries \citep{lamb1985quasi,wang2016brief}. Low-frequency variability in the mHz range was reported early in the black hole candidate Cyg X-1 \citep{frontera1975evidence}. Subsequently, using \textit{EXOSAT} observations, QPOs at frequencies of $\sim$20--40 Hz were discovered in GX~5$-$1 in 1985, marking one of the earliest detections of QPOs in accreting NS systems \citep{van1985intensity}. Notably, the presence of mHz QPOs in X-ray pulsars is crucial for unraveling the complex physics of neutron star environments and their accretion processes, and more than 20 additional NS X-ray binaries have been found to exhibit similar features \citep{james2010discovery}. The mHz QPOs appear as narrow peaks in power density spectra (PDSs) in the low-frequency range from $\sim$ 1 mHz to $\sim$ 1 Hz, with properties that depend on the source state, mass accretion rate, and other intrinsic physical parameters \citep{bachetti2010qpo}. In PDSs, they can be described by a Lorentzian function
\begin{equation}
P_{\nu}=\frac{A_0w}{\left( \nu -\upsilon _0 \right) ^2+\left( \frac{w}{2} \right) ^2}.   
\end{equation}
where $\upsilon _0$ refers to the central frequency of the signal, while $w$ denotes the full width at half maximum (FWHM), which quantifies the width of the peak. The amplitude of the signal is represented by $A_0$. The QPO signal is typically described by three parameters: the centroid frequency ($\upsilon _0$), the Q-factor, and the rms. The Q-factor is defined as $Q=\frac{\nu _0}{\omega}$, and it reflects the signal’s coherence. Typically, signals with $Q>2$ are identified as QPOs, whereas those with $Q<2$ are referred to as noise. The rms, which varies with the source flux, provides a measure of the signal’s strength and is proportional to the square root of the integrated power of its contribution to the PDS \citep{wang2016brief}.
\par
\citet{revnivtsev2001new} first discovered mHz QPOs in the NS-LMXBs 4U~1608--52, 4U~1636--53, and Aql~X-1, which appear only within a narrow luminosity range of, $L_{2\text{--}20\,\mathrm{keV}} = (5\text{-}11)\times10^{36}\,\mathrm{erg\,s^{-1}}$ and at frequencies of 7--9 mHz, showing strong flux variations below 5 keV. The mHz QPO in IGR J00291+5934 appears at $\sim$8 mHz when the persistent luminosity of the source reached 10-50 percent of the $L_{\rm Edd}$, and is most prominent below 3 keV \citep{ferrigno2017discovery}. IGR J17480-2446 exhibited mHz QPOs at 2.8-4.2 mHz emerging as faint and frequent thermonuclear bursts evolved smoothly as the persistent luminosity increased from about 0.1 to 0.5 $L_{\rm Edd}$ \citep{linares2012millihertz}. SAX J1808.4-3658 exhibits flares with characteristic rise and decay times of 0.2-1 s, producing strong QPOs at $\sim 1-5$ Hz \citep{patruno20091}. 4U 1730-22 exhibited QPOs at $\sim$ 4.5-8.1 mHz with $\sim$2\% rms, detected during a soft spectral state in multiple NICER observations \citep{mancuso2023detection}. These QPOs are thought to be associated with nuclear burning processes occurring on the NS surface. \citet{revnivtsev2001new} suggested that they arise from the surface burning operating at a certain range of mass accretion rates, while \citet{heger2007millihertz} proposed that they originate from marginally stable helium burning on the surface of accreting neutron stars, producing oscillations with a characteristic timescale of $\sim$100 s. Therefore, mHz QPOs represent a critical observational probe into the nuclear and accretion physics of LMXBs.
\par 
In HMXBs, mHz QPOs are often detected in the frequency range of approximately 1–1000 mHz, corresponding to characteristic timescales of one to hundreds of seconds \citep{devasia2011timing}. QPOs have been reported in about a dozen out of about 100 known HMXBs including 4U 0115+63 (10 mHz, 22 mHz, 41 mHz and 62 mHz \citep{ding2021qpos}), KS 1947+300 (20 mHz \citep{james2010discovery}), IGR J19294+1816 (30 mHz \citep{raman2021astrosat}), Cen X-3 (40 mHz \citep{liu2022detection}), SAX J2103.5+4545 (44 mHz \citep{inam2004discovery}), 1A 0535+262 (50 mHz \citep{finger1996quasi}), V 0332+53 (51 mHz \citep{takeshima1994discovery}), XTE J1858+034 (110 mHz \citep{paul1998quasi}), XTE J0111.2-7317 (1.27 Hz \citep{kaur2007quasi,devasia2011rxte}), Her X-1 (5 mHz, 10 mHz \cite{yang2025observations}). These oscillations are generally associated with accretion onto strongly magnetized NS and display a wide variety of temporal behaviors and energy-dependent characteristics. Their occurrence may be linked to inhomogeneities in the accretion flow, such as quasi-periodic structures in the accretion disk or plasma instabilities generated around the magnetospheric boundary \citep{van1987intensity,alpar1985gx5}. However, QPOs can also arise from other mechanisms. For example, the  0.2–0.5 Hz QPOs observed in RX J0440.9+4431 occur during the right wing of the pulse profile and are likely associated with hard X-ray flares \citep{li2024broad,malacaria2024discovery}. The multiple QPOs in 4U 0115+63 might be caused by instabilities in swirling flows, which are influenced by factors such as viscosity and magnetic fields \citep{ding2021qpos}. Therefore, studying QPOs in HMXBs is crucial for understanding the accretion processes and the physical conditions around strongly magnetized neutron stars.
\par
In this paper, we provide a systematic review of the observational characteristics and theoretical interpretations of mHz QPOs in accreting X-ray pulsars. We first discuss the current mainstream theoretical models in Section \ref{Theoretical Explanation}, including neutron star surface nuclear burning, Keplerian frequency model, beat frequency model, disk precession model, and orbital clump model. In section \ref{Observations of mHz QPOs in LMXBs} and \ref{Observations of mHz QPOs in HMXBs} , we present a comprehensive review of known mHz QPO X-ray pulsars and their characteristics in LMXBs and HMXBs, respectively, summarizing their observational properties and exploring their physical origins. Finally, we discuss current challenges and future research directions to facilitate a deeper understanding of the nature and physical origin of mHz QPOs in section \ref{Summary and prospective}.
\section{Physics Origin and Theoretical Models}
\label{Theoretical Explanation}
\par
The extensive survey of mHz QPOs presented in this work reveals a fundamental discrepancy between LMXBs and HMXBs, strongly suggesting distinct underlying physical mechanisms. In LMXBs, the collective evidence points towards a thermonuclear origin on the neutron star surface, whereas in HMXBs, the QPOs are predominantly governed by accretion flow instabilities and magnetospheric interactions. Other mechanisms, such as the propeller regime where a rapidly rotating magnetosphere can episodically inhibit and release accretion, have also been proposed but are less commonly invoked \citep{romanova2004propeller}. Accordingly, this section begins by summarizing the major theories and models for QPOs in both LMXBs and HMXBs. In Table \ref{tab:mHzQPO_models}, we present the brief description and summary of the known QPO production mechanisms, characteristics and the corresponding applicable sources. In the following, we mainly discuss five popular models for mHz QPOs in accreting neutron stars: nuclear burning, Keplerian frequency model, beat frequency model, disk precession and orbiting blob model. 
\par
\subsection{Nuclear burning on the neutron star surface}
\par
The nuclear burning model on the surface of neutron stars stands as the leading theoretical framework for explaining a specific class of mHz QPOs observed in LMXBs. This model was principally developed to explain the discovery of stable $\sim$7-9 mHz oscillations in sources like 4U 1636-536 and 4U 1608-52, which exhibited distinctive characteristics inconsistent with accretion driven variability. These key observational features, including their confinement to a narrow luminosity range, soft X-ray spectrum, and consistent disappearance following Type I X-ray bursts collectively pointed toward an origin in nuclear processes on the neutron star surface. The model posits that these oscillations arise from marginally stable helium burning in the neutron star's envelope. Within a specific range of mass accretion rates (typically $~0.5-1.5\times 10^{37} \rm erg\ s^{-1}$, or approximately 1-10\% of the Eddington luminosity). In this state, the nuclear burning process enters a critical regime where it is neither completely stable nor fully unstable. Instead of producing Type-I bursts or steady silent burning, the system settles into a quasi-steady oscillatory mode. The characteristic $\sim$100 s timescale of these oscillations corresponds directly to the thermal timescale of the burning layer, representing the natural period for the fuel to heat, ignite partially, cool, and subsequently re-accumulate toward another ignition episode \citep{revnivtsev2001new,bildsten1993rings}. The subsequent theoretical work by \citet{heger2007millihertz} provided comprehensive numerical validation through both simplified one-zone models and detailed multizone hydrodynamical simulations with the KEPLER code \citep{weaver1978presupernova}. Their work established that when the local accretion rate reaches a critical value of $\dot{m}\approx 0.924\ \dot{M}_{Edd}$, nuclear burning enters a marginally stable regime where the temperature dependence of heating and cooling rates nearly cancel, producing sustained oscillations. The characteristic period $P_{osc} \approx 4.44\times\sqrt{{t_{thermal} \times t_{accr}}}$ s naturally explains the observed $\sim$2-minute QPOs, where $t_{\rm thermal}$ is the thermal timescale of the burning layer, indicating how quickly the layer can heat up and cool down, while $t_{\rm accr}$ is the accretion timescale, representing the time required to accumulate enough fuel for nuclear ignition. Their models successfully reproduce the narrow accretion rate window and few-percent flux variations observed. To reconcile the order-of-magnitude discrepancy between the near-Eddington local accretion rate required by theory and the much lower globally observed rate, the authors proposed that the accreted fuel is confined to only about ten percent of the neutron star surface. This concentration elevates the local mass flux to the necessary level for marginally stable burning. Furthermore, they demonstrated that the resulting oscillation period depends sensitively on the surface gravity and the hydrogen fraction of the accreted material, thereby establishing these mHz QPOs as a unique probe for neutron star parameters \citep{heger2007millihertz}.

\begin{table*}
\centering
\caption{Summary of proposed models for mHz QPOs in neutron-star X-ray binaries.}
\label{tab:mHzQPO_models}
\begin{threeparttable}
\resizebox{\textwidth}{!}{
\begin{tabular}{lccc}
\hline\hline
Model & Applicable systems & Key physical ingredients & Ref.\tnote{a} \\
\hline
Nuclear burning 
& NS-LMXBs 
& Marginally stable nuclear burning on neutron-star surface 
& [1,2] \\

heartbeat-like instability 
&  IGR J00291+5931
& Accretion-driven hard–soft state cycles 
& [3] \\

KFM
& Accreting pulsars; \(\nu_{\rm QPO} > \nu_{\rm s}\)
& Keplerian motion at the inner disk 
& [4] \\

BFM 
& Accreting pulsars; \(\nu_{\rm QPO} < \nu_{\rm s}\) 
& Beat-frequency modulation at the inner disk
& [5] \\

MDPM 
& Strongly magnetized X-ray pulsars 
& Magnetically driven precession of a warped inner disk
& [6] \\

Orbiting Blob Model
& Pulsars with low-frequency QPOs and sidebands 
& Amplitude modulation by a single orbiting blob
& [7] \\

Instability near the $r_{co}$ of the disk
& Strongly magnetized X-ray pulsars 
& Centrifugal barrier–driven cyclic accretion near $r_{\rm co}$
& [8] \\

multi-vortex model
& Strongly magnetized X-ray pulsars 
& Polygonal vortices in unstable inner disks 
& [9] \\

X-ray flare
& Strongly magnetized X-ray pulsars 
& Transient polar-cap accretion flares 
& [10] \\
\hline
\end{tabular}
}
\begin{tablenotes} 
\footnotesize
\item[a] \parbox[t]{0.92\textwidth}{
References: 
[1] \citet{revnivtsev2001new}, 
[2] \citet{bildsten1993rings}, 
[3] \citet{belloni2000model}, 
[4] \citet{van1987intensity};
[5] \citet{alpar1985gx5}, 
[6] \citet{shirakawa2002precession}, 
[7] \citet{kommers1998sidebands};
[8] \citet{liu2022detection}, 
[9] \citet{ding2021qpos},
[10] \citet{li2024broad}
}
\end{tablenotes}
\end{threeparttable}
\end{table*}

\begin{figure}[h]
    \centering
    \includegraphics[width=1\textwidth]{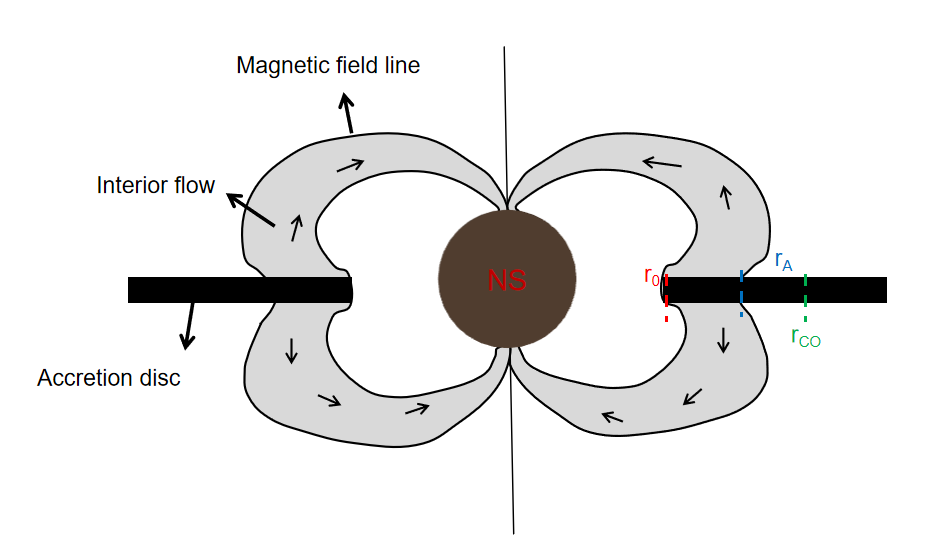}
    \caption{Schematic illustration of accretion onto a strongly magnetized neutron star (e.g., $B\sim 10^{12}$ G). The central black circle represents the neutron star, with magnetic field lines shown in black curves. Matter from the accretion disc (thick horizontal lines) is funneled along magnetic field lines, forming interior flows (shaded regions) toward the magnetic poles. The inner disc radius $r_0$, the Alfvén radius $r_A$, and the corotation radius $r_{co}$ are labeled. Arrows indicate the direction of plasma motion along the field lines.}
    \label{fig:fig3}
\end{figure}
\par
\subsection{Keplerian Frequency Model (KFM)}
\par
While the above model provides compelling explanations for mHz QPOs in LMXBs, where they are generally interpreted as signatures of marginally stable nuclear burning on the neutron star surface and whose disappearance following Type I X-ray bursts is attributed to the resetting of thermonuclear bursts resetting the burning conditions. In contrast to LMXBs, Type I X-ray outbursts observed in HMXBs are thought to be close to the periastron passage of the neutron star \citep{stella1986intermittent}, while Type II X-ray outbursts are possibly caused by the enhanced episodic outflow of the Be star \citep{paul2011transient}. Consequently, their connection to mHz QPOs differs fundamentally from that in LMXBs. As a result, the physical origin of QPOs in HMXBs would be governed by a distinctly different set of accretion dynamics which is uncertain, consequently the models are quite diverse at present. In Figure \ref{fig:fig3}, the accretion process of HMXBs and its key physical parameters including the corotation radius \(r_{\rm co}\), the Alfvén radius \(r_{\rm A}\), and the inner disk radius \(r_0\) are all labeled, providing a clear visual representation of the system’s accretion structure. Building upon the concept of accretion flow instabilities but proposing a radically different geometric mechanism, \citet{van1987intensity} proposed the Keplerian Frequency Model which was developed to explain the complex QPO behavior in the HMXBs. In the KFM, QPOs observed in the X-ray flux of accreting neutron stars are interpreted as the direct consequence of inhomogeneities or 'blobs' orbiting at the Keplerian frequency at the inner edge of the accretion disk ($r_0$). These inhomogeneities modulate the X-ray emission as they revolve around the neutron star, producing a periodic or quasi-periodic signal corresponding to their orbital motion. The QPO frequency therefore reflects the Keplerian orbital motion of matter near the disk magnetosphere boundary, and is given by

\begin{equation}
\nu_{\mathrm{QPO}} = \nu_{K} = \frac{1}{2\pi} \sqrt{\frac{GM}{r_0^3}},
\end{equation}
where \(M\) is the neutron star mass and \(r_0\) is the radius of the inner accretion disk. Physically, the inner disk radius \(r_0\) is determined by the balance between the magnetic pressure of the neutron star's dipole field and the ram pressure of the accreting plasma. This radius typically lies near the magnetospheric (Alfvén) radius \(r_{\rm A}\), so that 
\begin{equation}
r_0 \approx r_{\rm A}.
\end{equation}
The Alfvén radius \(r_A\) can be estimated as
\begin{equation}
r_{\rm A} \approx 2.3 \times 10^8 \, M_{1.4}^{1/7} R_6^{10/7} L_{37}^{-2/7} B_{12}^{4/7}~{\rm cm}.
\end{equation}
where $B_{12}$ represents the surface magnetic field strength of the neutron star in units of $10^{12}$ G, $M_{1.4}$ is the mass of the neutron star in units of 1.4 solar masses, $R_6$ is neutron star radius in units of $10^6$ cm, $L_{37}$ is accretion luminosity in units of $10^{37}$ erg/s.
\par
The KFM assumes that accretion onto the neutron star is continuous and that the QPOs trace the orbital dynamics of matter in nearly Keplerian motion without strong magnetic coupling. As the accretion rate increases, the magnetospheric boundary (and thus \(r_0\)) moves inward, leading to a higher Keplerian frequency and a corresponding increase in the QPO frequency. This model is particularly applicable to systems in which the observed QPO frequency exceeds the neutron star's spin frequency, implying that the disk material rotates faster than the stellar surface. In such cases, the QPO frequency provides a diagnostic of the inner disk radius and can be used to infer magnetic field strength, mass accretion rate, and the structure of the accretion flow near the neutron star.
\par
\subsection{Beat Frequency Model (BFM)}
\par
In the BFM proposed by \citet{alpar1985gx5}, QPOs observed in accreting neutron star systems are explained by the interaction between the Keplerian motion of matter at the inner edge of the accretion disk and the rotation of the neutron star's magnetic field. Blobs of matter orbit the neutron star at approximately the Keplerian frequency \(\nu_K(r_0)\) at the inner disk radius \(r_0\), while the neutron star rotates at the spin frequency \(\nu_{\rm s}\). 
The observed QPO frequency arises from the beat between these two motions:
\begin{equation}
\nu_{\rm QPO} = \nu_K(r_0) - \nu_{\rm s}.
\end{equation}
Additionally, the corotation radius \(r_{\rm co}\) is defined as the radius where the Keplerian frequency equals the neutron star spin frequency. If \(r_0 > r_{\rm co}\), the disk material rotates more slowly than the star and can be centrifugally inhibited from accreting, while for \(r_0 < r_{\rm co}\), matter can flow onto the neutron star surface. The BFM successfully explains QPOs with frequencies below the neutron star spin frequency, in contrast to the Keplerian Frequency Model (KFM), which requires \(\nu_{\rm QPO} > \nu_{\rm s}\). This makes BFM particularly relevant for slowly rotating X-ray pulsars or systems where the inner disk is truncated near or beyond the corotation radius.
\par
\subsection{Magnetic Disk Precession Model (MDPM)}
\par
The Magnetic Disk Precession Model explains low frequency QPOs in accretion powered X-ray pulsars through the warping and precession of the inner accretion disk caused by magnetic torques which was proposed by \citet{shirakawa2002precession}. In this model, the neutron star's magnetic field exerts a torque on the disk, inducing a precessional motion of the inner disk region. The characteristic precession period $\tau_{\rm prec}$ is given by
\begin{equation}
\tau_{\rm prec} = 776 \, \alpha^{0.85} \, L_{37}^{-0.71}~{\rm s}.
\end{equation}
where $\alpha$ is the dimensionless viscosity parameter of the accretion disk, and $L_{37}$ is the X-ray luminosity in units of $10^{37}~{\rm erg~s^{-1}}$. The corresponding QPO frequency is then
\begin{equation}
\nu_{\rm QPO} = \frac{1}{\tau_{\rm prec}}.
\end{equation}
This model is particularly applicable for HMXBs with slowly rotating neutron stars, where the inner disk can be significantly warped by magnetic torques. The precession leads to periodic modulations in the X-ray flux, naturally producing mHz QPOs. The predicted QPO frequency depends on both the disk viscosity and the accretion luminosity, providing a diagnostic of the inner accretion disk dynamics.
\par
\begin{figure}[h]
    \centering
    \includegraphics[width=0.8\textwidth]{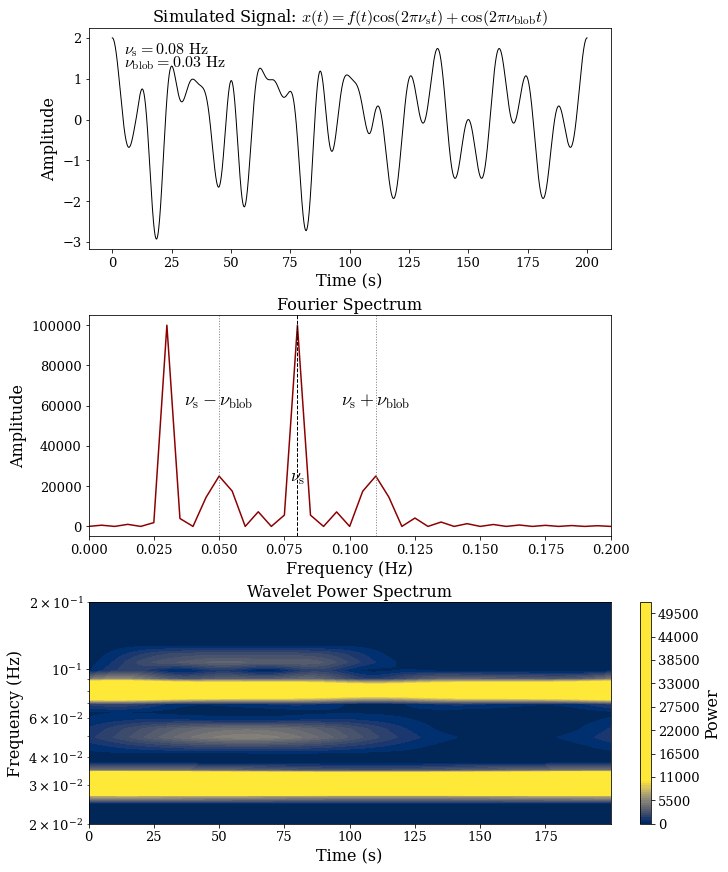}
    \caption{Simulated signal and its spectral analysis illustrating transient amplitude modulation due to an orbiting structure that absorbs and scatters X-rays from the pulsar beam.
    \textbf{(Top)} The time-domain signal 
    \( x(t) = f(t)\cos(2\pi\nu_{\rm s}t) + \cos(2\pi\nu_{\rm blob}t) \), 
    where the spin frequency \(\nu_{\rm s} = 0.08~{\rm Hz}\) and the modulation frequency \(\nu_{\rm blob} = 0.03~{\rm Hz}\). 
    The amplitude modulation function \(f(t)\) is applied only during a limited interval from \(t = 10\) to \(110~{\rm s}\), representing the transient interaction. 
    \textbf{(Middle)} The Fourier spectrum shows the main spin frequency peak and two sidebands at \(\nu_{\rm s} \pm \nu_{\rm blob}\), characteristic of amplitude modulation. 
    \textbf{(Bottom)} The wavelet power spectrum reveals that the modulation is temporally localized, consistent with the transient influence of a single orbiting structure. 
    }
    \label{fig:fig4}
\end{figure}
\subsection{Orbiting Blob Model}
\par
The Orbiting Blob Model, first proposed by \citet{kommers1998sidebands} for 4U 1626-67 and later found applicable to similar systems, provides a unified physical framework to explain the observed temporal phenomena in X-ray pulsars: a low-frequency QPO, symmetric sidebands around the spin frequency $\nu_{s}$. The model's core postulate is a single, coherent structure  "blob" orbiting the neutron star at a frequency $\nu_{blob}$. For a simple sine wave $\cos \left( 2\pi \nu_s \right)$, if the amplitude of this sine wave $f\left( t \right)$ changes over time with frequency $\nu_{blob}$, the Fourier transform will no longer have a single "spike" but will also have additional frequency components around the central frequency $\nu_{s}\pm \nu_{blob}$ forming sidebands. Simultaneously, as the blob moves along its orbit, it absorbs and scatters X-rays from the pulsar beam out of the line of sight. This periodic removal of flux from the direct beam is itself the source of the direct QPO signal at $\nu_{\text{blob}}$. In this refined view, the symmetric sidebands and the central QPO are not produced by separate mechanisms but are two direct consequences of the same amplitude modulation event. The model's elegance lies in attributing this complex power spectrum to a single, orbiting structure that modulates the observed intensity. The primary remaining puzzle is the physical origin and stability of the blob itself against shear forces in the accretion disk. To illustrate this mechanism, we constructed a numerical simulation in which the observed flux is represented as a sinusoidal pulsation whose amplitude is periodically modulated by a low frequency component, mimicking the effect of an orbiting blob crossing the line of sight. The simulated signal is written as
\begin{equation}
x(t) = f(t)\cos(2\pi\nu_{s}t) + \cos(2\pi\nu_{\text{blob}}t),
\end{equation}
where the modulation function $f(t)$ describes a transient amplitude variation over time. In our model, $f(t)$ is unity everywhere except during a finite interval (from $t=10$ to $t=110$~s), within which it is defined as
\begin{equation}
f(t) = 1 + \sin\!\left[ 2\pi(0.03~\text{Hz})(t - 10) \right].
\end{equation}
This represents an amplitude modulation at frequency $0.03~\text{Hz}$, corresponding to the orbital frequency of the blob. Physically, it models the periodic absorption or scattering of the pulsar’s X-ray beam as the blob passes through the line of sight. The Fourier transform of $x(t)$ exhibits distinct sidebands at $\nu_{s} \pm \nu_{\text{blob}}$, a direct consequence of this amplitude modulation, while the wavelet transform reveals that the modulation is temporally localized consistent with the transient influence of a single orbiting structure. The simulated process and its corresponding power spectra and wavelet power spectrum are illustrated in Figure \ref{fig:fig4}.

\section{Observations of mHz QPOs in LMXBs}
\label{Observations of mHz QPOs in LMXBs}
A proposed explanation for the mHz QPOs is their association with nuclear burning on the neutron star surface, with a predicted oscillation time scale of $\sim$100 s, as supported by observations of several LMXBs \citep{revnivtsev2001new,heger2007millihertz}. Nevertheless, this interpretation still remains difficult to confirm, since mHz QPOs do not always appear alongside type I X-ray bursts \citep{ferrigno2017discovery}. Observationally, mHz QPOs have been detected in several well-studied LMXBs, including 4U 1636-53, 4U 1608-52, Aql X-1, IGR J00291+5934, IGR J17480-2446, SAX J1808.4-3658, 4U 1730-22, and 4U 1626-67 \citep{ferrigno2017discovery,linares2012millihertz,patruno20091,mancuso2023detection,kommers1998sidebands,sharma2025sidebands}. For these sources, multiple observations spanning years with RXTE \citep{jahoda1996orbit}, XMM-Newton \citep{jansen2001xmm}, NICER \citep{gendreau2012neutron}, EXOSAT  \citep{taylor1981exosat}, Insight-HXMT  \citep{zhang2020overview} and INTEGRAL \citep{jensen2003integral} have provided detailed measurements of QPO centroid frequencies, amplitudes, energy dependence, and timing relative to type-I bursts. Summarizing the observational properties of these sources helps to constrain theoretical models and improves our understanding of the physical origin of mHz QPOs in neutron star LMXBs. Table \ref{tab:mHzQPO_summary} also provides a summary of these sources and their main observational properties.
\begin{table*}
\centering
\caption{Summary of mHz QPO properties in NS-LMXBs.}
\label{tab:mHzQPO_summary}
\begin{threeparttable}
\resizebox{\textwidth}{!}{
\begin{tabular}{lccccccc}
\hline\hline
Source & Distance & Spin & $P_{\rm orb}$ & $\nu_{\rm QPO}$ & rms & Model & Ref.\tnote{a} \\
 & (kpc) & (ms) & (hr) & (mHz) & (\%) &  &  \\
\hline
4U 1636--53 & $\sim$6 & 1.7 & 3.8 & 6--10 & 0.7--0.9 & nuclear burning & [1,2,3,4] \\
4U 1608--52 & 2.9--4.5 & 1.6 & 12.9 & 6--11 & 1.2--1.9 & nuclear burning & [1,5,6] \\
Aql X--1 & 4.5--6 & 1.8 & 18.95 & 6--9 & $\sim$2--4 & nuclear burning & [1,6] \\
IGR J00291+5934 & $\sim$4 & 1.6 & $\sim$2.5 & $\sim$8 & $\sim$12 & heartbeat-like instability\tnote{c} & [7] \\
IGR J17480--2446 (T5X2) & $\sim$6.3 & 90.54 & 21.27 & 2.8--4.2 & 1.3--2.2 & nuclear burning & [8,9,10] \\
SAX J1808.4--3658 & 3.5 & 2.4 & 2.01 & $\sim$1000 (1 Hz) & 10--125 & nuclear burning & [11] \\
4U 1730--22 & 6.9 & 1.7 & $<$2 & 4.5--8.1 & $\sim$2 & nuclear burning & [12] \\
4U 1626--67 & 5--13 & 7692 & 0.7 & 36--49, 83 & 20 & Orbiting Blob & [13,14] \\
\hline
\end{tabular}
}
\begin{tablenotes}
\footnotesize
\item[a] \parbox[t]{0.92\textwidth}{
References: 
[1] \citet{revnivtsev2001new}, 
[2] \citet{lyu2014discovery}, 
[3] \citet{lyu2015spectral}, 
[4] \citet{fei2021harmonic};
[5] \citet{yu2002kilohertz}, 
[6] \citet{mancuso2021drifts}, 
[7] \citet{ferrigno2017discovery};
[8] \citet{linares2012millihertz}, 
[9] \citet{linares2011cooling},
[10] \citet{papitto2011spin},
[11] \citet{patruno20091}, 
[12] \citet{mancuso2023detection},
[13] \citet{kaur2008study}, 
[14] \citet{sharma2025sidebands}
}
\item[b] Only the fundamental mHz QPO frequency is listed; harmonics are not shown.
\item[c] \parbox[t]{0.92\textwidth}{The heartbeat-like instability refers to QPOs caused by transitions between a hard state with an obscured inner disk and two soft states with fully observable disks at different temperatures, driven by cyclic changes in local accretion rate \citep{belloni2000model}}
\end{tablenotes}
\end{threeparttable}
\end{table*}

\par
\begin{figure}[h]
    \centering
    \includegraphics[width=0.35\textwidth]{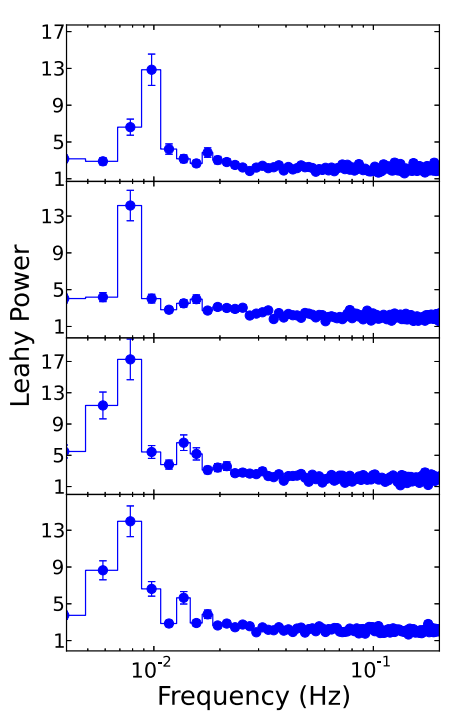}
    \includegraphics[width=0.45\textwidth]{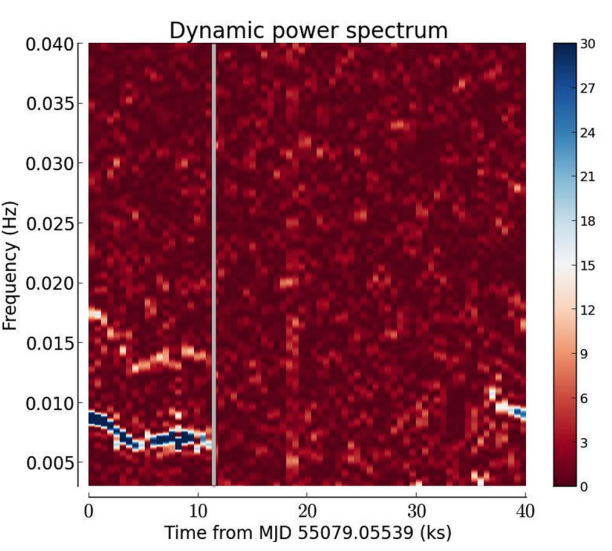}
    \caption{(Left) Average power spectra of 4U 1636-53 from four XMM-Newton observations (top to bottom) in the 0.2-5 keV adapted from \citep{lyu2015spectral}. (Right) A clear mHz QPO around 8-10 mHz is visible in each observation, accompanied by a weaker second harmonic. Dynamic power spectrum of 4U 1636-53 from XMM-Newton observations taken from \citep{lyu2014discovery}. The grey vertical line marks the occurrence of an X-ray burst.}
    \label{fig:fig5}
\end{figure}
\begin{figure}[h]
    \centering
    \includegraphics[width=0.8\textwidth]{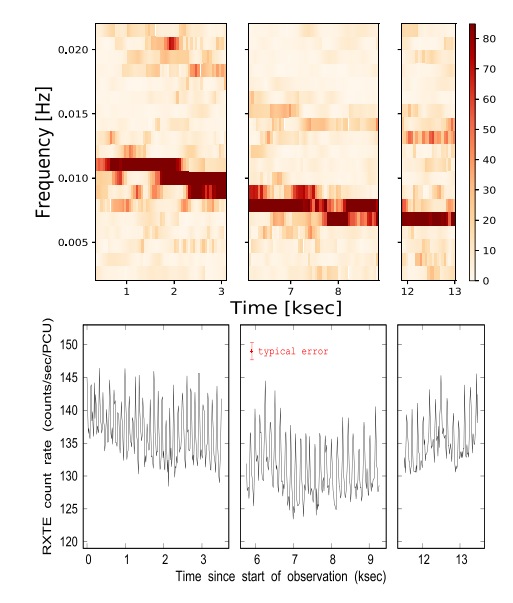}
    \caption{(Top) Dynamic power spectrum of 4U 1608-52 from RXTE observation. The mHz QPO frequency gradually drifts from approximately 10-11 mHz down to 6-7 mHz, with a weak harmonic occasionally visible. (Bottom) Background-subtracted RXTE/PCA light curve in the 2-5 keV. The figure is taken from \citep{mancuso2021drifts}.}
    \label{fig:fig6}
\end{figure}

\subsection{4U 1636-53}
4U 1636-53 is a NS-LMXB located approximately 6 kpc away, containing a 0.1-0.25 $M_{\odot}$ companion star \citep{giles2002burst}. The system has an orbital period of about 3.8 h \citep{pedersen1982simultaneous} and hosts a neutron star with a spin period of $\sim$1.72 ms \citep{zhang1997u,strohmayer2002evidence}. The 4U 1636-536 was first observed to exhibit weak mHz QPOs at a stable frequency of $\sim$7-9 mHz, with an rms of 0.7-0.9\% in the 0.8-5 keV, consistently detected in RXTE and EXOSAT observations spanning 1985–1999 \citep{revnivtsev2001new}. \citet{altamirano2008millihertz} reported that in 4U 1636-53, the mHz QPO frequency gradually declines over time and vanishes before the appearance of a type I X-ray burst. \citet{lyu2014discovery} tracked the possible evolution of the mHz QPO by computing a highly oversampled dynamic power spectrum using overlapping intervals and the Lomb–Scargle periodogram method \citep{lomb1976least} and revealed a 7-8 mHz QPO with a possible 14-16 mHz harmonic, present only before and after the type I X-ray burst. Based on an analysis of X-ray timing data from 2007-2009 using power density spectra and Lomb-Scargle periodograms, \citet{lyu2015spectral} discovered the mHz QPOs exhibited significant frequency variations, drifting between approximately 6 and 10.4 mHz. They also found that these QPOs consistently disappear before type-I X-ray bursts. Spectral analysis showed no correlation between QPO frequency and neutron star surface temperature. The average power spectra from the four XMM-Newton observations reveal a significant QPO at about 8-10 mHz with a weak second harmonic in each observation and the dynamic power spectra illustrate the frequency evolution of the mHz QPOs and their consistent disappearance before X-ray bursts as shown in Figure \ref{fig:fig5}. Using 2–5 keV X-ray light curves from RXTE PCA, \citet{fei2021harmonic} find that the rms amplitude ratio of the harmonics to the fundamental remains constant ($\sim$60\%). As such, 4U 1636-53 is an ideal source for understanding the relation between different manifestations of nuclear burning, providing key insights into marginally stable burning and linking the thermal response of the LMXB envelope to the occurrence of type I X-ray bursts.
\par
\subsection{4U 1608-52}
4U 1608-52 is a transient LMXB, first identified in 1972 \citep{belian1976discovery,grindlay1976uhuru} and classified as an atoll source based on its spectral and timing characteristics \citep{hasinger1989two}. The system exhibits recurrent outbursts every $\sim$85 days to 1-2 years and harbors a rapidly spinning neutron star with a spin frequency of 619 Hz \citep{galloway2008thermonuclear}. Its orbital period is approximately 0.537 days \citep{wachter2002closer}, with a system inclination estimated between $30^\circ$ and $40^\circ$ \citep{degenaar2015nustar} and a source distance ranging from 2.9 to 4.5 kpc \citep{galloway2008thermonuclear}. A variety of thermonuclear burst behaviors have been observed from 4U 1608-52, including single, multipeaked, and superburst events \citep{penninx1989exosat}. Based on RXTE/PCA observations during the March 1996 and March 1998 outbursts, 4U 1608-52 exhibited mHz QPOs with a centroid frequency of $\sim$7.5 mHz, a width of $\sim$1.7-2.6 mHz, and an rms of $\sim$1.2-1.9\% in the 2-5 keV. These oscillations were detected only when the source luminosity was in the range of $(0.5\text{--}1.1) \times 10^{37} \ \mathrm{erg/s}$ and showed a strong energy dependence, with the rms amplitude decreasing significantly above 5 keV and becoming undetectable above $\sim$7 keV. \citet{yu2002kilohertz} discovered that in 4U 1608–52, the frequency of the kHz QPOs anti-correlates with the 2–5 keV X-ray flux modulated by the 7.5 mHz QPOs. This anti-correlation supports the scenario in which the increased radiation from the neutron star surface during the mHz QPO cycle exerts outward pressure, pushing the inner accretion disk to a larger radius. The QPOs often exhibit systematic downward frequency drifts, with rates between $\sim 0.28$ and $1.9~\mathrm{mHz~ks^{-1}}$ over timescales of a few kiloseconds, as well as smaller stochastic variations below $\sim 2~\mathrm{mHz}$ as illustrated in Figure \ref{fig:fig6}. The mHz QPOs disappear during Type I X-ray bursts, with rms amplitudes dropping from $\sim 2$--$3\%$ to $<0.5\%$, and show no clear overall correlation with the soft (2--5 keV) X-ray count rate \citep{mancuso2021drifts}. Studying these QPOs is therefore crucial for understanding the link between nuclear burning and inner disk processes.

\subsection{Aql X-1}
Aql X-1 is a transient LMXB discovered in 1965 \citep{friedman1967distribution}, exhibiting regular outbursts with recurrence times of roughly 125-300 days \citep{priedhorsky1984long,campana2013mining,kitamoto1993unstable}. Distance estimates range from 4.5 to 6 kpc \citep{jonker2004distances,mata2017donor}, with an orbital period of $\sim$18.95 hours \citep{chevalier1991discovery} and an inclination of $36^\circ$-$47^\circ$ \citep{mata2017donor}. Spectral and timing studies classify Aql X-1 as an atoll-type source \citep{hasinger1989two}. Additionally, coherent millisecond X-ray pulsations at 550.27 Hz have been detected \citep{casella2008discovery}. Based on the reanalysis of RXTE data, Aql X-1 exhibited mHz QPOs with a centroid frequency of $\sim$6-7 mHz that disappeared after the onset of a type I X-ray burst. These oscillations showed a soft energy spectrum, becoming undetectable above $\sim$7 keV \citep{revnivtsev2001new}. The mHz QPOs occasionally exhibit downward frequency drifts, with one clear case showing a decrease from $\sim 9.0$ to $\sim 6.0$~mHz over $\sim 3.2$~ks, corresponding to an average drift rate of $\sim 0.92$~mHz~ks$^{-1}$. Additional downward drifts of $\lesssim 1.3$~mHz generally associated with the softest spectral states were observed in several instances, mostly at frequencies below 9~mHz, occurring in both the banana branch (BB) and lower island state (IS) of the CCD. These behaviors are similar to those in 4U~1608--52 and 4U~1636--53 observations \citep{mancuso2021drifts}.
\subsection{IGR J00291+5934}
IGR J00291+5934 is an accreting millisecond X-ray pulsar (AMXP), a type of LMXB with a spin period of 1.67~ms \citep{markwardt2004igr}. Systems like this typically have orbital periods ranging from $\sim$40 minutes to a few hours and spend most of their lifetime in a quiescent state, with a typical X-ray luminosity of $10^{31}$–$10^{32}$~erg~s$^{-1}$. Occasionally, they undergo X-ray outbursts lasting from weeks to months, during which accretion onto the NS dramatically increases, reaching luminosities up to $\sim 10^{36}$–$10^{38}$~erg~s$^{-1}$ \citep{patruno2020accreting}. The first QPO at $\sim$8 mHz with rms $\sim$ 12 per cent was discovered in the IGR J00291+5934 during its 2015 outburst. This QPO shows a direct connection to the source's flaring activity and exhibits strong energy dependence, being predominantly detected below 3 keV while modulating both the persistent flux and pulsed emission. These observational characteristics suggest the QPO likely originates from a heartbeat-like accretion instability \citep{belloni2000model} rather than thermonuclear burning processes \citep{ferrigno2017discovery}.
\subsection{IGR J17480-2446}
\begin{figure}[h]
    \centering
    \includegraphics[width=0.8\textwidth]{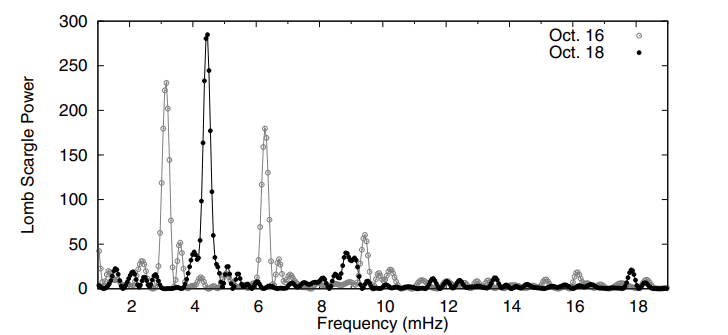}
    \caption{Lomb-Scargle periodograms of the mHz QPOs observed on 2010 October 16 (gray) and 2010 October 18 (black) for T5X2 taken from \citep{linares2012millihertz}. The periodograms clearly display the QPO harmonic structure, with up to four harmonic structure, and show the variation in $\nu_{QPO}$ between the two observations.}
    \label{fig:fig7}
\end{figure}

IGR J17480-2446 (hereafter T5X2) is a transient LMXB located in the globular cluster Terzan~5. It was discovered in October 2010 by INTEGRAL \citep{chenevez2010further} and quickly monitored with RXTE, which revealed 11 Hz X-ray pulsations and burst oscillations at the same frequency \citep{strohmayer2010exo,altamirano2010transient}. The orbital period of T5X2 is 21.27454(8) hr, and the NS magnetic field is estimated to lie in the range $10^8$--$10^{10}$~G \citep{motta2011x,miller2011fast}. Near the outburst peak, T5X2 exhibited spectral and timing properties typical of Z sources, when accreting at roughly half the Eddington rate \citep{altamirano2010transient}. The mHz QPOs observed in IGR J17480-2446 occur at frequencies of 2.8-4.2 mHz, significantly lower than typical mHz QPOs from other neutron star systems. As shown in Figure \ref{fig:fig7}, these QPOs exhibit both detectable frequency variations within this range and a strong harmonic structure with multiple overtones. They display remarkable coherence with $Q \geq 3$ and rms of 1.3\%-2.2\% in the 2-60 keV. Unlike other known mHz QPO sources, their amplitude increases with energy up to $\sim$20 keV. These QPOs demonstrate a smooth evolutionary connection with thermonuclear bursts, they gradually transform into burst sequences as the accretion rate decreases, and providing the clear observational evidence linking mHz QPOs to marginally stable nuclear burning \citep{linares2012millihertz}.
\subsection{SAX J1808.4-3658}
\begin{figure}[h]
    \centering
    \includegraphics[width=0.8\textwidth]{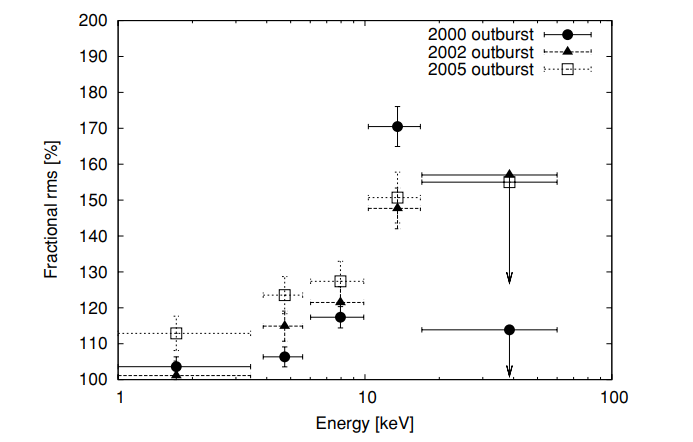}
    \caption{Energy dependence of the rms of the 1 Hz QPO for SAX J1808.4-3658 taken from \citep{patruno20091}. For all three outbursts, the rms amplitude increases with photon energy, shown here up to 17 keV where the count rates are significant. In the 17–60 keV range, only upper limits (at 98\% confidence) could be determined for the three outbursts.}
    \label{fig:fig8}
\end{figure}

SAX J1808.4-3658 is a transient LMXB with an orbital period of $P_{\rm orb} = 2.01$~hr and a very low-mass companion star of $\sim$0.05--0.08~M$_\odot$ discovered by BeppoSAX in 1996 \citep{zand1998discovery} and later identified as the AMXP in 1998 with the RXTE \citep{wijnands1998millisecond}. Since then, five major outbursts have been recorded: 1996, 1998, 2000, 2002, 2005, and 2008, all showing remarkably similar profiles with durations of several weeks and recurrence intervals of about 2.5 years. Each outburst features a fast rise (2--5 days), a short high-flux peak, a slow decay ($\sim$10 days), and a rapid decline (3--5 days), followed by a low-flux re-flaring state lasting for weeks to months \citep{campana2008swift,wijnands2001erratic,wijnands2003xmm,campana2008swift,wijnands2004observational}. During all outbursts, 401 Hz pulsations were detected even at low luminosities ($\sim10^{34}$ erg s$^{-1}$). The 1 Hz QPO in SAX J1808.4-3658 exhibits a centroid frequency ranging from 0.8 to 1.6 Hz that correlates positively with X-ray flux, while its rms varies dramatically between 10\% and 125\% as shown in Figure \ref{fig:fig8}. This modulation displays Q < 2 with occasional harmonic structure. The QPO emerges near the end of the outburst fast decay phase and recurs sporadically during re-flares \citep{patruno20091}.
\subsection{4U 1730-22}
4U 1730–22 is a transient LMXB first discovered in 1972 by the Uhuru satellite \citep{cominsky1978transient}. The system underwent an outburst that lasted about 200 days before entering a quiescent phase \citep{chen1997properties,cominsky1978transient}. After this, it was not observed to flare again for nearly 50 years, until June 2021 \citep{kobayashi2021maxi}. In July 2021, the source underwent a rapid brightening, and the NICER satellite detected a thermonuclear X-ray burst for the first time, solidifying the identification of the compact object as a neutron star \citep{bult2021nicer}. Distance estimates derived from photospheric radius expansion bursts during the 2022 outburst placed the source at 6.9 $\pm$ 0.2 kpc \citep{bult2022thermonuclear}. Additionally, spin period at 1.71 ms was observed in the 2022 outburst  \citep{li2022discovery}. The mHz QPOs detected in the 4U 1730-22 exhibited characteristic frequencies between approximately 4.5 and 8.1 mHz, with rms amplitude of 2\%. These oscillations were exclusively observed when the source was in a soft spectral state, as confirmed by X-ray color analysis, and within luminosity range of $\left( 0.5-1.2 \right) \times 10^{37}\ erg\ s^{-1}$ in the 2-20 keV. A key identifying feature was their connection to thermonuclear activity, as they were found to immediately precede type-I X-ray bursts and subsequently vanish once a burst occurred. Furthermore, the rms of the QPOs in some observations showed a tendency to increase with energy up to around 3 keV, consistent with theoretical predictions that link such low frequency oscillations to thermal instabilities in the accreted fuel layer near the transition between stable and unstable burning regimes \citep{mancuso2023detection}. 
\subsection{4U 1626–67}
\begin{figure}[h]
    \centering
    \includegraphics[width=0.8\textwidth]{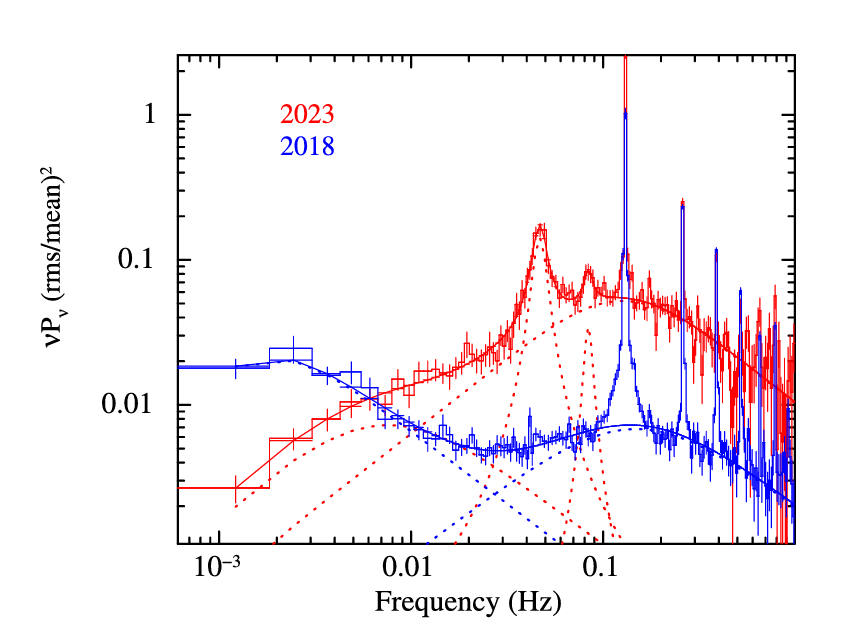}
    \caption{The 3-30 keV power density spectra of 4U 1626-67 from AstroSat observations in 2018 (spin-up torque state) and 2023 (spin-down torque state), modeled with multiple Lorentzian components taken from \citep{sharma2025sidebands}. Sharp peaks associated with the neutron star spin period (7.66 s) and its harmonics were excluded from the fit.}
    \label{fig:fig9}
\end{figure}

4U 1626-67 is a LMXB and a well-known accreting X-ray pulsar, discovered in 1972 \citep{giacconi1972uhuru}. It is characterized by an extremely short binary orbit of approximately 42 minutes and a spin period of about 7.7 s \citep{rappaport1977discovery,middleditch19804u1626}. This source is most notable for its unique behavior of undergoing long-term torque reversals, cycling between steady spin-up and spin-down states over decades, which are accompanied by significant changes in its X-ray luminosity, pulse profile, and energy spectrum \citep{chakrabarty1997torque,camero2009new,sharma20234u}. The QPOs in 4U 1626-67 exhibit a strong dependence on the neutron star's torque state. During spin-up phases, they are weak or undetectable, as seen with the faint 0.04 Hz QPO in the first spin-up epoch and their general absence in the second \citep{shinoda1990discovery}. Prominent $\sim$48 mHz QPOs consistently emerge during spin-down states, reported by multiple observatories \citep{owens1997complex,chakrabarty1998high,krauss2007high}. The evolution of these QPOs across different torque states is directly illustrated in Figure \ref{fig:fig9}. Furthermore, the QPO centroid frequency shows long-term evolution, increasing from $\sim$36 mHz to $\sim$49 mHz over a decade of spin-up and subsequently decreasing at a rate of $\sim$0.2 mHz/yr during the following spin-down phase, highlighting a dynamic link between the QPO phenomenon and the pulsar's accretion torque \citep{kaur2008study}. A $\sim$46 mHz QPO is detected during the current spin-down episode. This fundamental QPO exhibits a strong positive correlation between its rms and energy. Furthermore, the QPO modulates the coherent pulsations, generating symmetric sidebands in the PDS. This is evidenced by the detection of a lower sideband at $\sim$83 mHz ($\nu_{s}$ - $\nu_{QPO}$) and an upper sideband at $\sim$177 mHz ($\nu_{s}$ + $\nu_{QPO}$) \citep{kommers1998sidebands,sharma2025sidebands}. The amplitude of these sidebands, particularly the upper one, shows a clear energy dependence, being significantly stronger in the 15-30 keV. The origin of the QPOs can be well explained by the Orbiting Blob Model. In this scenario, as the blob periodically absorbs or scatters X-rays from the pulsar beam, it modulates the amplitude of the pulsations, producing both the central QPO at the blob’s orbital frequency and sidebands around the spin frequency. Therefore, the detection of sidebands in 4U 1626-67 confirms that this QPO physically modulates the amplitude of the pulsed X-ray emission \citep{kommers1998sidebands,sharma2025sidebands}.

\begin{table*} 
\centering 
\caption{Summary of mHz QPO properties in NS-HMXBs.} 
\label{tab:mHzQPO_HMXBs} 
\begin{threeparttable} 
\resizebox{\textwidth}{!}{ 
\begin{tabular}{lcccccc} 
\hline\hline 
Source & Distance & Spin Period & Orbital Period & QPO Frequency & Model & Ref.\tnote{a} \\
& (kpc) & (s) & (days) & (mHz) & & \\
\hline
Cen X-3 & 6.9 & 4.8 & 2.087 & 33--43 & Instability near $r_{\rm co}$ of the disk\tnote{b}. & [1,2,3] \\
Her X-1 & 6.6 & 1.24 & 1.7 & 5-9,10 & BFM$\&$ MDPM & [4] \\
IGR J19294+1816 & 11 & 12.5 & 117.2 & 30, 51, 114 & Orbiting Blob & [5] \\
4U 0115+63 & 7 & 3.6 & 24.3 & 1, 2, 5-14.9, 31-62.5 & Multi-vortex model\tnote{c} & [6,7] \\
KS 1947+300 & 10 & 18.7 & 40.415 & 21.5 & - & [8] \\
SAX J2103.5+4545 & 6.5 & 358 & 12.68 & 44 & KFM & [9] \\
1A 0535+262 & 2.13 & 104 & 111 & 30--70 & - & [10] \\
V 0332+53 & 7 & 4.39 & 34 & 51 & - & [11] \\
XTE J1858+034 & 10 & 222 & 81 & 196 & BFM & [12,13] \\
XTE J0111.2-7317 & 60 & 31 & 31.4 & 1266 & KFM$\&$ BFM & [14] \\
RX J0440.9+4431 & 2.44 & 205 & 150 & 200--500 & X-ray flare\tnote{d} & [15] \\
4U 1901+03 & 12.4 & 2.76 & 22.58 & 135 & BFM & [16] \\
SMC X-1 & 60 & 0.72 & 3.9 & 10 & -\tnote{e} & [17] \\
\hline 
\end{tabular}
} 
\begin{tablenotes} 
\footnotesize 
\item[a] \parbox[t]{0.95\textwidth}{References: 
[1] \citet{liu2022detection}, 
[2] \citet{van2021new}, 
[3] \citet{liu2023measurements},
[4] \citet{yang2025observations}, 
[5] \citet{Yang_2025}, 
[6] \citet{roy2019laxpc}, 
[7] \citet{ding2021qpos}, 
[8] \citet{james2010discovery}, 
[9] \citet{reig2004discovery}, 
[10] \citet{finger1996quasi}, 
[11] \citet{qu2005discovery}, 
[12] \citet{manikantan2024energy}, 
[13] \citet{mukherjee2006variable}, 
[14] \citet{kaur2007quasi}, 
[15] \citet{li2024broad}, 
[16] \citet{james2011flares}, 
[17] \citet{angelini1989qpos} 
} 
\item[b] \parbox[t]{0.95\textwidth}{The instability near the $r_{\rm co}$ arises because the centrifugal barrier prevents accretion when the inner disk radius is outside $r_{\rm co}$. As a result, matter accumulates in the inner disk until the increased pressure overcomes the barrier, pushing the disk inward, which triggers an accretion episode and initiates a new cycle \citep{liu2022detection}.}

\item[c] \parbox[t]{0.95\textwidth}{The multi-vortex model proposes that the inner disk edge can form a meta-stable polygonal vortex structure under the combined effects of magnetic and centrifugal forces. Changes in disk parameters lead to repeated disruption and reconstruction, naturally explaining the transient sub-Hz QPOs. Different QPO timescales correspond to polygon mode switching and rotation of sub-vortices \citep{ding2021qpos}.} 
\item[d] \parbox[t]{0.95\textwidth}{The X-ray flare model suggests that the QPOs originate from transient flares at the right wing of the pulse profile, caused by short-term injections of high-energy material onto the neutron star’s polar cap, producing quasi-periodic modulation only in the hard X-ray band \citep{li2024broad}.} 
\item[e] \parbox[t]{0.95\textwidth}{"-" represents no adequate model for the source.}
\end{tablenotes} 
\end{threeparttable} 
\end{table*}

\section{Observations of mHz QPOs in HMXBs}
\label{Observations of mHz QPOs in HMXBs}
Although mHz QPOs discovered in LMXBs have been extensively studied and are relatively well explained within the nuclear burning on the neutron star surface, the understanding of HMXBs remains considerably incomplete. Compared to LMXBs, the observational data on mHz QPOs in NS-HMXBs are relatively scarce, and these sources exhibit complex diversity. Although a considerable number of QPO sources have been identified, their physical origins and dominant mechanisms remain undetermined, traditional models such as the BFM and the KFM face significant challenges in explaining the observed characteristics of several sources, such as energy dependence and frequency-luminosity relations. This section provides a systematic review of the observed properties of mHz QPOs in representative sources, which are summarized in Table \ref{tab:mHzQPO_HMXBs}. It is expected that this effort will provide crucial observational constraints for different theoretical models, thereby deepening the understanding of accretion processes in strong magnetic field environments.
\begin{figure}[h]
    \centering
    \includegraphics[width=0.8\textwidth]{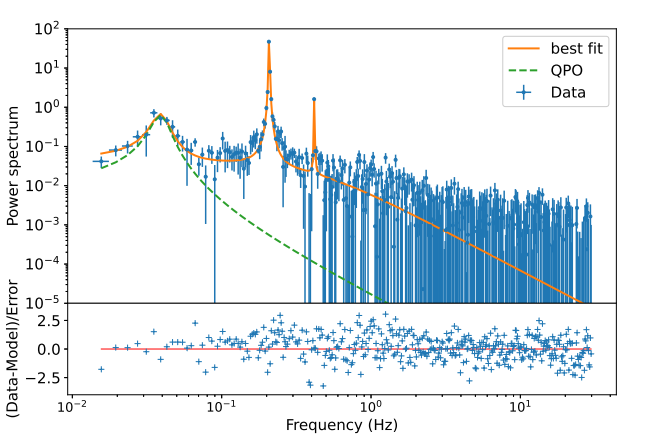}
    \caption{Power spectrum of X-ray curves for Cen X-3 observed by Insight-HXMT (ME: 10–30 keV), showing a clear QPO at $\sim$40 mHz (green dashed line) taken from \citep{liu2022detection}. The spectrum also includes the neutron star spin frequency ($\sim$0.2 Hz) and its harmonic ($\sim$0.4 Hz).}
    \label{fig:fig10}
\end{figure}
\subsection{Cen X-3}
Cen X-3 is a HMXB that was first discovered by \citet{chodil1967spectral}. It is the first confirmed X-ray pulsar, as identified by \citep{giacconi1971discovery}. The system consists of a NS that orbits an O-type supergiant star, V779 Cen \citep{schreier1972evidence}. The NS has a spin period of approximately 4.8 s \citep{giacconi1971discovery}, and there is a spin-up trend observed over time, indicating an increase in its rotation speed \citep{finger1994observations,tsunemi1996long}. The orbital period of the system is about 2.087 days \citep{falanga2015ephemeris}, with an inclination angle of around $79^\circ$ \citep{sanjurjo2021x}, meaning the system exhibits an eclipsing behavior that lasts roughly 22\% of the orbit \citep{chodil1967spectral}. This makes Cen X-3 an ideal system for studying eclipsing X-ray binaries. The neutron star in Cen X-3 has a mass of $1.34\,M_\odot$ \citep{van2007determination} with a strong magnetic field ranging from $\sim (2 - 3) \times 10^{12}$ G \citep{santangelo1998bepposax}. The optical companion, the O-type supergiant V779 Cen, has a mass of approximately $20.2\,M_\odot$ and a radius of about $12\,R_\odot$ \citep{van2007determination,naik2011x}. Early X-ray observations showed QPOs at approximately 35 mHz \citep{takeshima1991quasi}. Later, \citet{raichur2008quasi} reported QPOs with peak frequencies around 40 mHz and 90 mHz. The $\sim$40 mHz QPO in the Cen X-3 exhibited a central frequency ranging between $\sim$33 and 43 mHz, with a high rms averaging $\sim$9\% detected by Insight-HXMT in 2020 \citep{liu2022detection}. Based on the Insight-HXMT observations in 2020, the PDS of Cen X-3, when the $\sim$40 mHz QPO was detected, can be characterized in Figure \ref{fig:fig10}. Its frequency showed a systematic evolution with orbital phase, increasing from the lower during phases 0.1-0.4 to higher values at phases 0.4-0.8. The QPO was exclusively detected when the source was in a soft spectral state and displayed a clear energy dependence, with the rms amplitude decreasing as energy increased from 2 to 20 keV. These distinct characteristics, particularly the high rms and energy dependence, differentiate it from the mHz QPOs typically seen in LMXBs and point to an origin likely tied to instabilities at the inner edge of an accretion disk truncated near the corotation radius \citep{liu2022detection}.
\subsection{Her X-1}
\begin{figure}[h]
    \centering
    \includegraphics[width=0.8\textwidth]{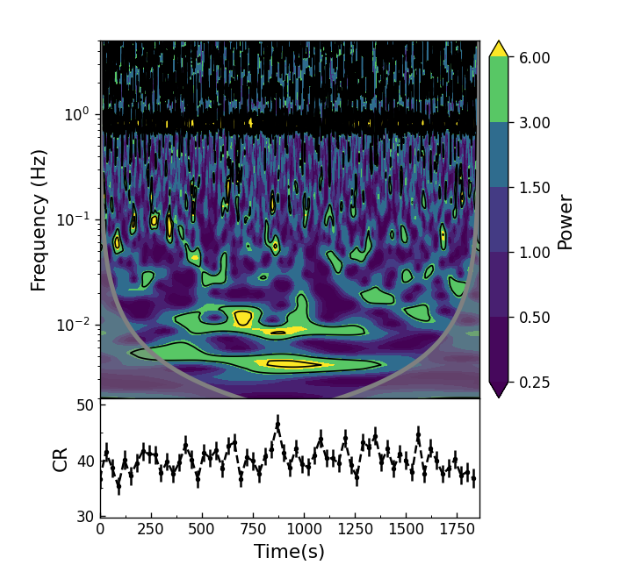}
    \caption{Wavelet analysis of the Her X-1 light curve in the 10–30 keV in August 2018, taken from \citep{yang2025observations}. The top panel displays the local wavelet power spectrum and the bottom panel shows the corresponding count rates sampled every 30 s.}
    \label{fig:fig11}
\end{figure}
Hercules X-1 (Her X-1), an eclipsing binary X-ray pulsar with a spin period of 1.24 s in a $\sim 1.7$ d circular orbit was first discovered in 1972 during Uhuru observations \citep{giacconi1973further}. Shortly after its discovery, the blue variable 13th-magnitude star HZ Her was identified as its optical companion \citep{davidsen1972identification}.  The Her X-1/HZ Her binary system exhibits a 35-day superorbital cycle showing two maxima in intensity: main-high (MH) state and short high (SH) state, with the MH lasting approximately 10 days and the SH with flux reaching about 30\% of the MH maximum lasting about 5 days, and separated by low states (LS) of roughly 10 days with flux at approximately 1\% of the maximum \citep{giacconi1973further,scott1999rossi,igna2011hercules}. Observational studies across multiple wavelengths have revealed mHz QPOs in Her X-1. Ultraviolet observations with the Hubble Space Telescope detected QPOs at frequencies of $8\pm 2$ and $43\pm 2$ mHz in the continuum emission, likely originating from the heated atmosphere of the companion star HZ Her \citep{boroson2000discovery}. Complementary optical observations using the Keck Observatory identified low-frequency QPOs centered around 35 mHz \citep{o2001keck}. Additionally, X-ray data from RXTE observations in 1996 and 1998 showed evidence of excess power at approximately 10 mHz, which may correspond to X-ray QPOs in this system \citep{moon2001discovery, o2001keck}. Based on Insight-HXMT observations, Her X-1 exhibits complex mHz QPOs characterized by the presence of two distinct types: relatively stable $\sim$10 mHz QPOs whose frequencies show no dependence on X-ray luminosity, and lower frequency $\sim$5-9 mHz QPOs whose centroid frequencies demonstrate a positive correlation with luminosity. Wavelet analysis reveals these QPO features are highly transient and exhibit rapid frequency evolution on timescales of hundreds of seconds with the resulting plots shown in Figure \ref{fig:fig11}. The concurrent detection of $\sim$10 mHz QPOs in both X-ray and UV bands suggests a common physical origin, likely related to a beat frequency mechanism at the inner accretion disk. In contrast, the luminosity-dependent $\sim$5 mHz QPOs may originate from magnetic disk precession \citep{yang2025observations}.
\subsection{IGR J19294+1816}
\begin{figure}[h]
    \centering
    \includegraphics[width=0.8\textwidth]{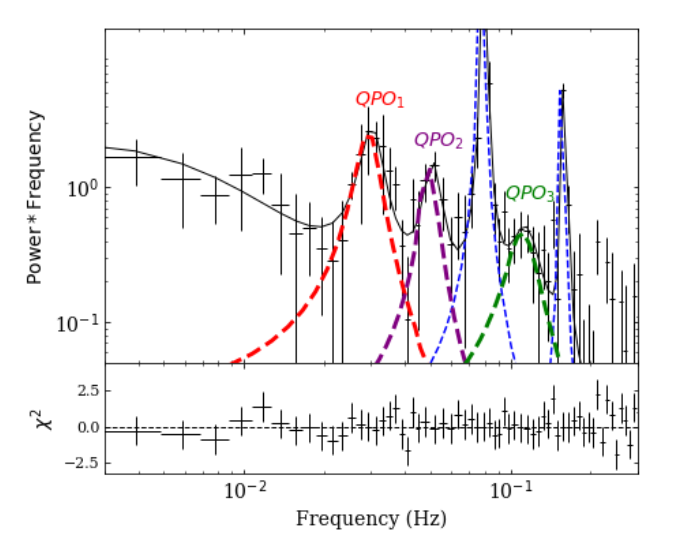}
    \caption{Power density spectrum for IGR J19294+1816 in the 25-50 keV taken from \citep{Yang_2025}. The neutron star’s 12.48 s spin peak and its harmonics are shown by blue dotted lines, while QPOs at $\sim$30 mHz, $\sim$50 mHz, and $\sim$110 mHz are marked by red, purple, and green dotted lines, respectively.}
    \label{fig:fig12}
\end{figure}
IGR J19294+1816 was discovered during an outburst in 2009 by the IBIS/ISGRI instrument on the INTEGRAL Gamma-ray Observatory \citep{turler2009integral}. This source located at a distance of 11 $\pm$ 1 kpc \citep{rodes2018igr}, was identified as a BeXB with a pulsation period of 12.4 seconds through Swift observations \citep{rodriguez2009nature}. Observations from the Swift/BAT monitor also revealed long-term flux variations with a periodicity of approximately 117.2 days \citep{2009ATel.2008....1C}. The source is notable for its cyclotron resonance scattering feature (CRSF) at around 40 keV, which indicates the presence of a strong magnetic field of approximately $4.6\times 10^{12}G$ \citep{tsygankov2019study, raman2021astrosat}. A significant timing feature of IGR J19294+1816 is the detection of QPOs at 0.032 $\pm$ 0.002 Hz, observed during the 2019 type I outburst, with an rms of about 18\%. These QPOs exhibit a positive correlation with energy, as shown by observations from AstroSat and XMM-Newton \citep{raman2021astrosat, manikantan2024energy}. Based on Insight-HXMT timing analysis, \citet{Yang_2025} reported the detection of multiple QPOs in the IGR J19294+1816. Their analysis revealed a fundamental QPO at $\sim$30 mHz, accompanied by a pair of sideband QPOs at approximately 51 mHz and 114 mHz as shown on Figure \ref{fig:fig12}. The sideband QPOs are consistent with a modulation of the neutron star's spin frequency ($\sim$80 mHz) by the fundamental QPO. The fundamental QPO detected in the 10-50 keV energy exhibits a strong energy dependence, with the rms increasing significantly with photon energy. Wavelet analysis indicated that all QPO features are transient. This discovery establishes IGR J19294+1816 as the second strongly-magnetized pulsar that exhibits such symmetric sideband signals modulating the pulsed emission, providing a key case study for understanding the underlying physical mechanisms.
\subsection{4U 0115+63}
\begin{figure}[h]
    \centering
    \includegraphics[width=0.8\textwidth]{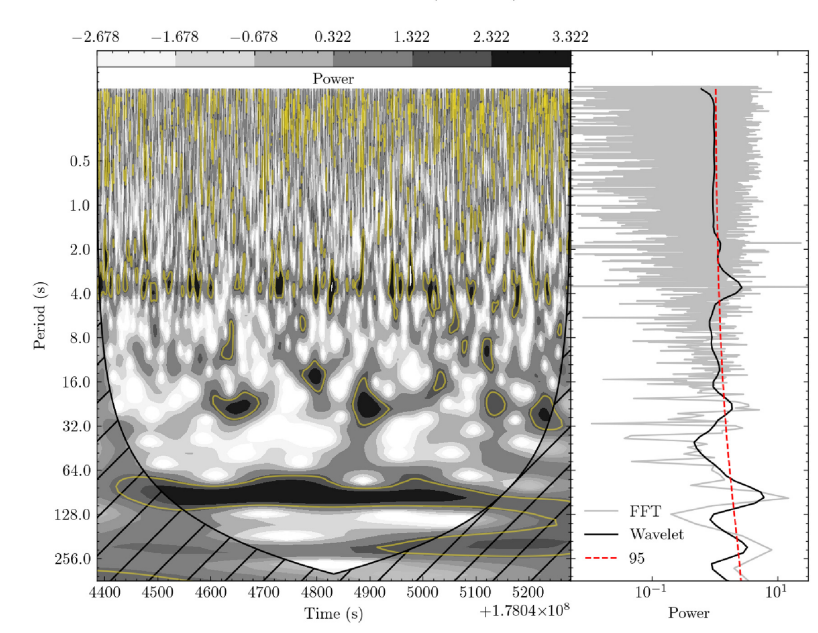}
    \caption{Wavelet power map for 4U 0115+63 in the 30-50 keV taken from \citep{ding2021qpos}. The left panel shows the wavelet power spectrum with the 95\% confidence level indicated by yellow lines. The right panel displays the corresponding global wavelet spectrum. Most QPO features have disappeared, except for the 100 s and 22 s oscillations.}
    \label{fig:fig13}
\end{figure}

The source 4U 0115+63 is a BeXB which was initially discovered by the Uhuru satellite \citep{giacconi1972uhuru}. The system comprises a neutron star and an O9e type companion star identified as V635 Cas, which has a visual magnitude of \( V \approx 15.5 \) \citep{cominsky1978discovery, rappaport1978orbital}. The distance to the system is estimated to be approximately 7 kpc \citep{negueruela2001x}. Based on its spectral type, the mass of the companion star V635 Cas is expected to be around \( 19\,M_{\odot} \) \citep{vacca1996lyman}. The neutron star is characterized by a spin period of \(\sim 3.6\) seconds \citep{cominsky1978discovery}. The binary orbit has an eccentricity of \(\sim 0.34\) and an orbital period of 24.3 days, as determined from early observations \citep{kelley1981search}. Multiple QPOs around 27-46 mHz were revealed with the RXTE during the source's 1999, 2004, and 2008 outbursts \citep{dugair2013detection}. Furthermore, an earlier RXTE study by \citet{heindl1999discovery} reported the presence of a $\sim$2 mHz QPO. This lower-frequency QPO was later confirmed, and an additional $\sim$1 mHz QPO was identified, through AstroSat/LAXPC observations taken during the 2015 outburst on October 24 \citep{roy2019laxpc}. 4U 0115+63 exhibited a new QPO at approximately 10 mHz ($\sim$100 s) was discovered based on Insight-HXMT observations of its 2017 outburst, alongside previously reported QPOs in the $\sim$16–32 s and $\sim$67–200 s periods \citep{ding2021qpos}. The stability (S-factor, defined as the ratio of the effective oscillation time to the total time \cite{chen2022waveletB,chen2024different}) and Q-factor of these QPOs showed a positive correlation with X-ray luminosity, while the QPO periods themselves showed no such dependence. Wavelet phase maps revealed that the lower-frequency QPOs ($\sim$67-200 s) frequently exhibited phase drifting, whereas the higher-frequency ones ($\sim$16-32 s) were relatively stable as shown in Figure \ref{fig:fig13}. The authors explain these QPOs as resulting from vortex structures forming at the magnetospheric boundary, where larger vortices generate the $\sim$100 s QPOs and smaller sub-vortices produce the $\sim$16-30 s QPOs, with phase drifts arising from transitions between different vortex patterns. 
\subsection{KS 1947+300}
KS 1947+300 was first discovered in June 1989 with the coded-mask imaging spectrometer aboard the Kvant module of the Mir Space Station \citep{skinner1989ks}. Pulsations with a period of 18.7 s were later detected by BATSE on board the Compton Gamma Ray Observatory (CGRO) in 1994 \citep{swank2000ks}. The optical counterpart of KS 1947+300 is a B0 Ve star with a visual magnitude of $V = 14.2$ and moderate reddening, indicating a distance of approximately 10 kpc \citep{negueruela2003x}. \citet{galloway2004frequency} derived the system’s orbital parameters: an orbital period of $P_{\mathrm{orb}} = 40.415 \pm 0.010$ days, a projected semi-major axis of $a_{\mathrm{x}}\sin i = 137 \pm 3$ lt-s, and an eccentricity of $e = 0.033 \pm 0.013$. Observational studies of the transient pulsar KS 1947+300 have revealed a QPO at 0.0215 Hz, characterized by high rms (15.4\%) and moderate coherence (Q $\sim$3.6) appeared exclusively during the late decline phase of the 2001 outburst and exhibited marginally increasing amplitude with energy below 20 keV. Theoretical frameworks struggle to account for these observational characteristics, as the QPO frequency falls below the neutron star's spin frequency, ruling out the Keplerian Frequency Model, while the Magnetospheric BFM yields inconsistent estimates for the inner-disk radius $(9.6\times10^{3}\,\mathrm{km})$ and the magnetospheric radius (\(2.76\times10^{4}\,\mathrm{km}\)), given the established magnetic field strength of \(B = 2.5\times10^{13}\,\mathrm{G}\) which was inferred from the correlation between the spin-up torque and the broad band X-ray luminosity of this source \citep{tsygankov2005observations}. The transient nature of this oscillation, coupled with the absence of cyclotron features, suggests that either previous magnetic field strength determinations require substantial revision or fundamentally different physical mechanisms must be invoked to explain such low-frequency variability in accreting pulsar systems \citep{james2010discovery}. 
\begin{figure}[h]
    \centering
    \includegraphics[width=0.4\textwidth]{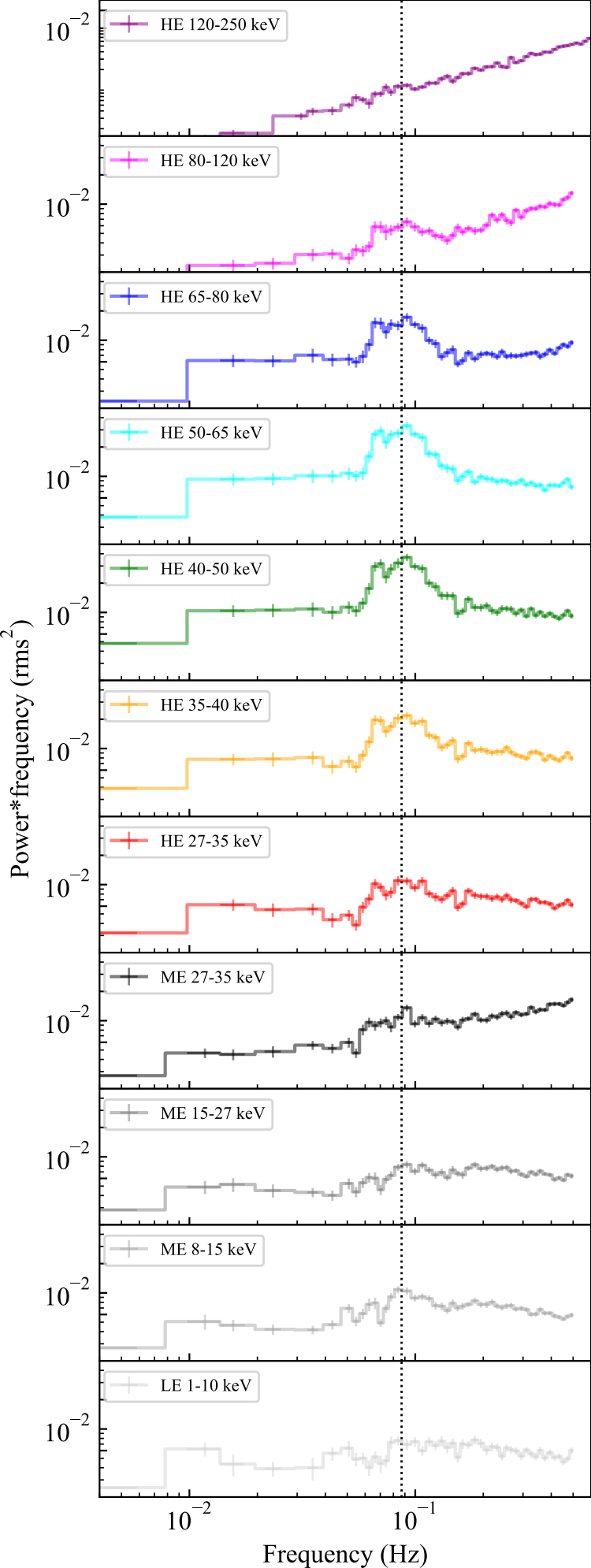}
    \caption{Power density spectra of X-ray light curves during the 2020 outburst for 1A 0535+262 in different energy bands taken from \citep{ma2022high}. The vertical dashed line marks the centroid frequency of the mHz QPO.}
    \label{fig:fig14}
\end{figure}
\subsection{SAX J2103.5+4545}
SAX J2103.5+4545 is a transient HMXB that was first discovered by the BeppoSAX during an outburst in 1997 \citep{hulleman1998discovery}. The system features a neutron star with a pulse period of approximately 358.6 s, orbiting its companion with a period of 12.68 days in an eccentric orbit (e$\sim$0.4) around its B-type companion star \citep{baykal2000discovery}. SAX J2103.5+4545 exhibits QPOs with a centroid frequency of 0.044 Hz (22.7 s period), detected in the 0.9-11 keV energy band during XMM-Newton EPIC-PN observations. The QPO feature displays an rms of 6.6\% and a Q-factor $\sim$ 7.2, indicating moderate coherence. These oscillations show a transient nature, being absent in simultaneous RXTE/PCA observations in the 3-20 keV, suggesting either strong energy dependence or intrinsic transient behavior. Theoretical interpretation suggests the QPO frequency corresponds to Keplerian motion at the inner accretion disk radius \citep{inam2004discovery}.
\subsection{1A 0535+26}
1A 0535+262 is a transient BeXB system discovered in 1975, consisting of a neutron star orbiting a Be-type companion star \citep{coe1975hard}. The system exhibits an orbital period of approximately 111 days and a neutron star spin period of about 104 s \citep{finger1996quasi}. Located at a distance of 2.13 kpc \citep{bailer2018estimating}, this source has undergone multiple giant outbursts since its discovery, with the 2020 outburst being the brightest recorded at approximately 11 Crab in the Swift/BAT band \citep{arabaci2020optical}. The source shows significant aperiodic variability in the 30-70 mHz, which was found to scale with both the neutron star’s spin-up rate and the pulsed X-ray flux, in agreement with the predictions of the BFM and KFM \citep{finger1996quasi}. However, subsequent analysis by \citet{camero2012x} revealed that the mHz QPO became more pronounced at higher energies but vanished below 25 keV. This pronounced energy dependence casts doubt on the applicability of both models to fully explain the observed behavior. Those properties have also been studied across multiple energy bands up to 120 keV using Insight-HXMT observations \citep{ma2022high}. These mHz QPOs display energy-dependent characteristics and have been detected at unprecedentedly high energies above 80 keV during the 2020 outburst as shown in Figure \ref{fig:fig14}, providing crucial insights into accretion processes in strongly magnetized neutron star systems \citep{ma2022high}.
\subsection{V 0332+53}
V0332+53 is a transient HMXB system, discovered during an outburst in 1973 with the Vela 5B observatory \citep{terrell19841973}. The system consists of an accreting neutron star and its optical counterpart, the O8-9 Ve star BQ Cam, located approximately 7 kpc \citep{bernacca1984identification, honeycutt1985spectrophotometry}. The neutron star has a pulse period of about 4.375 s and orbits its companion every 34 days in a moderately eccentric orbit (e $\sim$0.3) \citep{stella1985discovery}. Cyclotron resonance scattering features indicate a magnetic field of about $2.7\times 10^{12} G$ at the neutron star's surface. The QPOs exhibit a centroid frequency at approximately $(5.1 \pm 0.5) \times 10^{-2}$ Hz, corresponding to a timescale of about 20 s, with a full width at half maximum (FWHM) of $(3.3 \pm 1.5) \times 10^{-2}$ Hz and an rms amplitude of $4.8 \pm 1.2\%$ \citep{takeshima1994discovery}. Based on RXTE observations of the 2004/2005 outburst, the known 0.05 Hz QPO in V0332+53 was characterized by a stable centroid frequency of 0.049 $\pm$ 0.007 Hz, which showed no significant evolution even as the X-ray flux changed by a factor of $\sim$4.5. Its rms was energy-dependent, remaining roughly constant at 4-6\% below 10 keV but decreasing by approximately a factor of two at higher energies \citep{qu2005discovery}. The origin of the 0.05 Hz QPO in V0332+53 cannot be explained by the traditional magnetospheric BFM or the KFM, as both predict a correlation between QPO frequency and X-ray flux, which is not observed \citet{qu2005discovery}.
\subsection{XTE J1858+034}
\begin{figure}[h]
    \centering
    \includegraphics[width=0.8\textwidth]{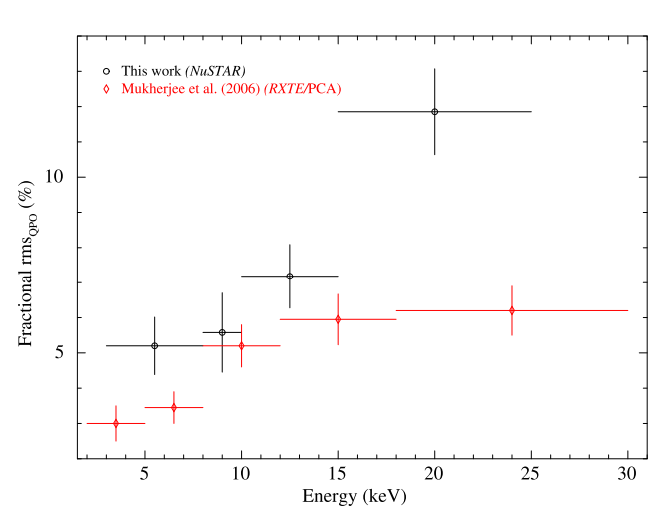}
    \caption{XTE J1858+034 QPO rms energy dependence on in November 2019 from NuSTAR (black; \citep{manikantan2024energy}) and RXTE/PCA in April-May 2004 (red; \citep{mukherjee2006variable}).}
    \label{fig:fig15}
\end{figure}

XTE J1858+034 was first discovered in February 1998 by the RXTE during a transient outburst \citep{bradt2001x} which was thought to be a transient accretion-powered X-ray pulsar in a Be/X-ray binary system \citep{paul1998quasi}. A recent study by \citet{tsygankov2021x} has posed a challenge to the conventional classification of this system, proposing an alternative interpretation that it may be a Symbiotic X-ray Binary with a K/M-type stellar companion. The source exhibits strong sinusoidal X-ray pulsations with a period of approximately 221 seconds \citep{takeshima1998xte,paul1998quasi} and has shown multiple transient outbursts, including events in 2004 detected by both INTEGRAL \citep{doroshenko2008study}. The 110 mHz QPO in this source, first identified by \citealt{paul1998quasi} using RXTE/PCA, was later found by \citet{mukherjee2006variable} to exhibit a variable centroid frequency (140-185 mHz) during subsequent outbursts, and its rms amplitude also revealed a strong correlation with photon energy. Based on the NuSTAR observations of the 2019 outburst, the QPO in XTE J1858+034, detected at a centroid frequency of $\sim$196 mHz, exhibits a strong positive correlation between its rms and photon energy, increasing from $\sim$5\% in the 3-8 keV to $\sim$11\% in the 15-25 keV as shown in Figure \ref{fig:fig15}. This energy-dependent rms trend, while consistent with the overall positive correlation seen in previous RXTE observations, was found to continue rising above 10 keV. QPO properties are best explained by the BFM, as the inner accretion disk radius derived from this model is consistent with the estimated $Alfv\acute{e}n$ radius \citep{manikantan2024energy}.

\subsection{XTE J0111.2-7317}
XTE J0111.2-7317 is a transient accretion-powered X-ray pulsar located in the direction of the Small Magellanic Cloud (SMC) with spin frequency of 0.032 Hz \citep{yokogawa2000asca}. It was first discovered in November 1998 by the RXTE and simultaneously detected in hard X-rays with BATSE on board the Compton Gamma Ray Observatory (CGRO) during an outburst with a flux ranging from 18 to 37 mCrab \citep{wilson1998xte}. Follow-up observations with ASCA confirmed coherent pulsations with an X-ray flux of $3\times 10^{-10}erg\ s^{-1}\ cm^{-2}$ in the 0.7-10 keV and revealed a distinct soft excess component, now interpreted as reprocessed emission from the inner accretion disk \citep{paul2001detection}. The pulsar exhibits a spin-up trend with a characteristic timescale of about 20 years \citep{yokogawa2000asca}. The optical counterpart has been identified as a B0.5–B1 Ve star \citep{covino2001discovery}. The QPO detected in the transient pulsar XTE J0111.2-7317 exhibited a high centroid frequency of 1266 $\pm$ 18 mHz, making it one of the highest-frequency QPOs observed in the HMXBs. This feature was remarkably narrow, with a width of 70 $\pm$ 10 mHz, and had an rms amplitude of 2.52\%.  Based on the observational properties, the study explores theoretical models to explain the QPO's origin. Both the KFM and the BFM are considered applicable because the QPO frequency is much higher than the neutron star's spin frequency. Both models point to an inner accretion disk radius of approximately $1.4\times 10^8$cm for the QPO production region. By equating this radius with the magnetospheric radius, the neutron star's magnetic field strength is estimated to be in the range of $(2.2-4.4)\times 10^{12}$ G. This provides a crucial indirect estimate of the magnetic field in the absence of a detected cyclotron absorption line in the spectrum \citep{kaur2007quasi}.
\subsection{RX J0440.9+4431}
\begin{figure}[h]
    \centering
    \includegraphics[width=1\textwidth]{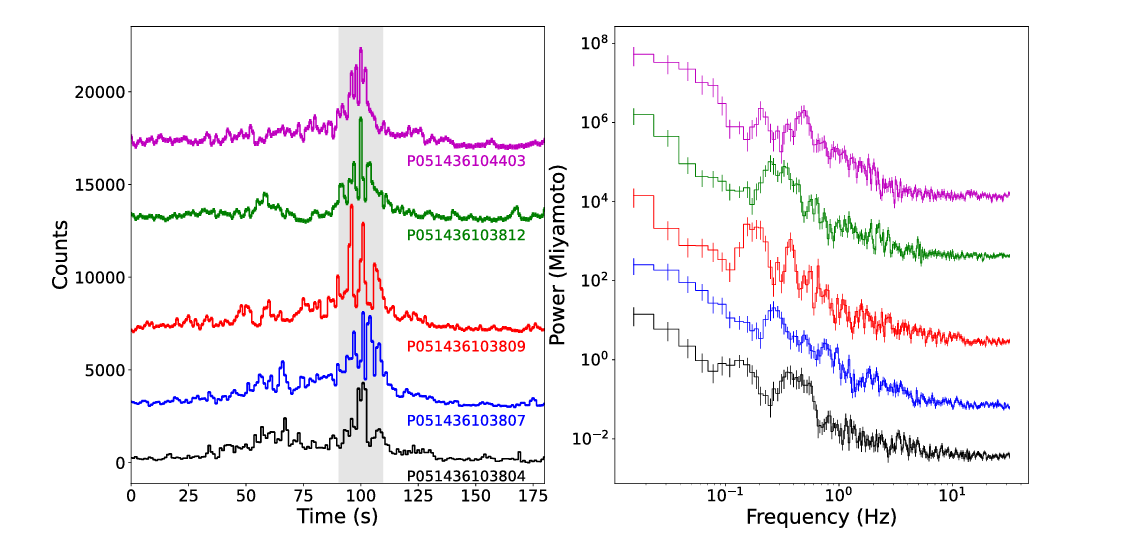}
    \caption{\textbf{(left)} Net light curves for five Insight-HXMT observations and \textbf{(right)} corresponding power density spectra in the 30-100 keV for RX J0440.9+4431 taken from \citep{li2024broad}. Each pulse and power is shown in a distinct color for clarity. QPOs are detected within short flares at the peaks of five individual pulses, indicated by grey shading.}
    \label{fig:fig16}
\end{figure}

RX J0440.9+4431 is an X-ray binary system consisting of a slowly rotating neutron star with a spin period of 205 seconds in orbit around a Be-type companion star \citep{motch1996new} and is characterized by outburst activity, first detected by MAXI in 2010 \citep{morii2010maxi}. RX J0440.9+4431 located at a distance of about 2.44 kpc \citep{bailer2021estimating}, its outburst luminosities are below typical Type-I outbursts in Be X-ray binaries \citep{reig2011x}. However, the source increased significant when it experienced its brightest recorded outburst at the end of 2022, reaching a peak flux of approximately 2.25 Crab \citep{coley2023continued}. This intense outburst enabled detailed studies across a wide luminosity range, leading to the determination of its critical luminosity at around $3\times 10^{37}\rm erg\ s^{-1}$ \citep{mandal2023probing}. The cyclotron absorption features and the phase-resolved spectra reveal a high surface magnetic field ($B \sim 10^{13}$ G) for the pulsar \cite{epili2025aa}. The QPO in RX J0440.9+4431 appears at approximately 0.2 Hz. It is strictly transient, detected only during the peak of the source's outburst when the luminosity surpasses a critical threshold of about $2.8\times 10^{37}\rm erg\ s^{-1}$. Furthermore, the QPO exhibits a strong spin-phase dependence, being visible only near the peak of the neutron star's pulse profile, which indicates its emission is likely linked to a specific periodically observable region like the base of an accretion column \citep{malacaria2024discovery}. Insight-HXMT has revealed that the $\sim$0.2-0.5 Hz QPOs originate from short and phase-dependent hard X-ray flares within individual pulsar pulses \citep{li2024broad} as directly shown in Figure \ref{fig:fig16}.
\subsection{4U 1901+03}
\begin{figure}[h]
    \centering
    \includegraphics[width=0.75\textwidth]{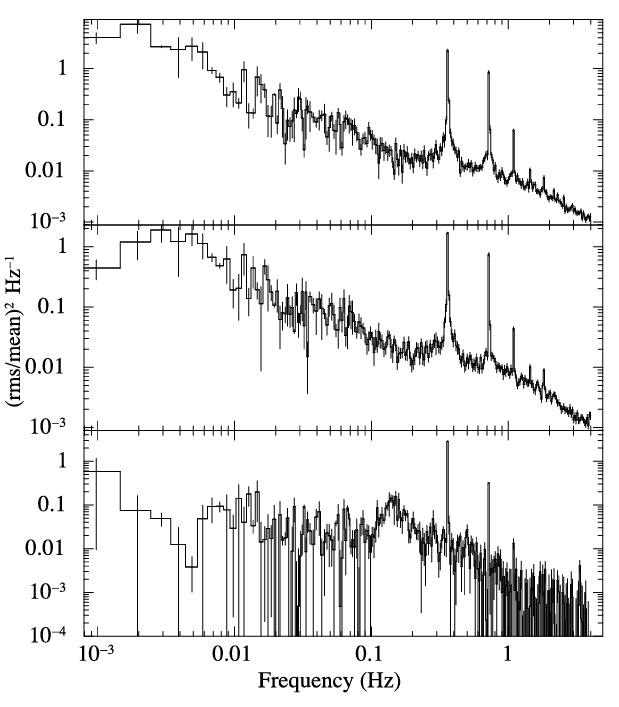}
    \caption{Power density spectra of the pulse peak with (top and middle panels) and without (bottom panel) broadening for 4U 1901+03 taken from \citep{james2011flares}. The QPO feature at $\sim$0.13-0.14 Hz can be clearly seen in the bottom panel.}
    \label{fig:fig17}
\end{figure}

The HMXB 4U 1901+03 was first discovered during its initial outburst in 1970-1971 by the Uhuru and Vela 5B satellites \citep{forman1976uhuru,priedhorsky1984long}. The source harbors an accreting neutron star characterized by a 2.763 s spin period with a 22.58 day low eccentricity (e$\sim$0.035) orbit, placing it in the peculiar class of wide orbit yet low eccentricity X-ray binaries \citep{galloway2005discovery}. Despite multi-wavelength efforts, no optical or infrared counterpart has been identified \citep{galloway2005discovery}. Observations across missions, including hard X-ray monitoring by INTEGRAL in 2003, have revealed that its emission properties are highly dynamic, showing luminosity-dependent pulse profiles and a strong pulse-phase dependence \citep{lei2009phase}. The QPO detected at $\sim$0.135 Hz during the low luminosity of its 2003 outburst as shown in Figure \ref{fig:fig17}, exhibited observational features including a high rms of 18\% and a restriction to energies below 11 keV. Its frequency being lower than the neutron star's spin frequency supports interpretation via the BFM, where the QPO originates from the interaction between the Keplerian accretion disk and the pulsar's magnetosphere, yielding a magnetic field estimate of $\sim0.3\times 10^{12}$G \citep{james2011flares}.
\subsection{SMC X-1}
SMC X-1 is a HMXB located in the Small Magellanic Cloud \citep{kallivayalil2006smc,lang1998studies}. It is in a 3.9 day binary system with a B0 supergiant companion star \citep{schreier1972discovery,angelini1991discovery}, which eclipses the X-ray source for approximately 0.6 days \citep{nagase1989accretion}. The neutron star itself has a remarkably stable spin period of 0.72 s \citep{price1971x} and is experiencing a steady spin-down \citep{angelini1991discovery}. Another remarkable feature of SMC X-1 is its extreme luminosity, which can reach up to $10^{39}\rm erg\ s^{-1}$, briefly exceeding the Eddington limit for a typical 1.4 solar mass neutron star by a factor of five \citep{price1971x}. The system also exhibits long-term flux variability on timescales of tens of days, potentially due to a precessing accretion disk \citep{hu2013superorbital,trowbridge2007tracking}. The QPOs in SMC X-1 were discovered from observations conducted in 1984 \citep{angelini1989qpos}. Analysis of the EXOSAT data revealed a primary QPO feature at a low frequency of approximately 0.01 Hz, which manifested as $\sim$100 s flaring activity in the light curves and exhibited a low coherence, broad hump in the power spectrum with an rms variation of about 8\%. A second, broader excess was also detected around $\sim$2 Hz. The properties of the 0.01 Hz QPO, including its presence and width, were found to be dependent on the source's persistent X-ray emission luminosity \citep{angelini1989qpos}.
\begin{figure}[h]
    \centering
    \includegraphics[width=0.8\textwidth]{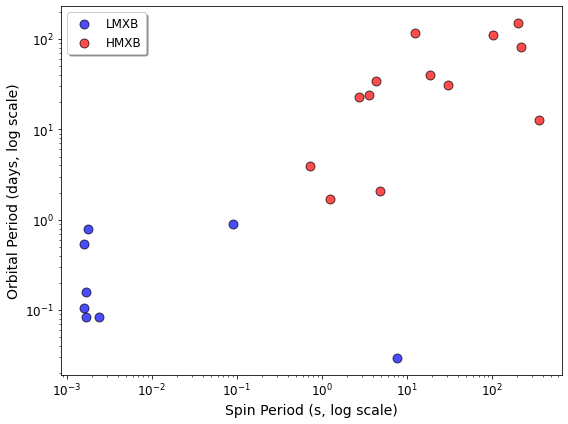}
    \caption{Corbet diagram of the orbital period versus spin period for X-ray pulsars: the HMXBs and LMXBs discussed in this work are marked with red and blue circles, respectively. The data are taken from the references listed in Tables 2 and 3. The X-ray pulsars of LMXBs and HMXBs have the different distribution in this diagram, and have different physical origins of mHz QPOs. In addition, a LMXB 4U 1626--67 has a spin period (7.7 s) much longer than other LMXBs, which also shows the different QPO features and physical origin compared with other LMXBs. }
    \label{fig:fig18}
\end{figure}
\section{Summary and prospective}
\label{Summary and prospective}
\par
Building upon the preceding discussion of QPOs in neutron star systems, the mHz QPOs observed in NS-LMXBs are now widely interpreted as manifestations of marginally stable nuclear burning on the neutron star surface. This interpretation is supported by a coherent set of observational characteristics that appear consistently across multiple sources. These QPOs are detected only within a narrow luminosity range, typically around 1–10\% of the Eddington limit, precisely where theory predicts the burning to become marginally stable due to its sensitivity to the local accretion rate, as shown in 4U 1608–52 and 4U 1636–53. Their oscillation period, about 100 seconds, corresponds to the thermal timescale of the neutron star’s burning layer, reflecting the natural cycle of fuel accumulation, ignition, and cooling. A strong piece of evidence for their thermonuclear nature lies in their disappearance immediately following Type-I X-ray bursts, a behavior observed in sources like 4U 1636–53 and Aql X-1, indicating that the bursts exhaust or disrupt the fuel layer responsible for the oscillations. Additionally, these QPOs exhibit a pronounced energy dependence, their rms peaks in the soft X-ray band below $\sim$5 keV and vanish at higher energies, consistent with thermal emission from the neutron star surface, as demonstrated in 4U 1730-22. The most compelling observational link between marginally stable and unstable burning is provided by IGR J17480–2446, where mHz QPOs were seen to evolve smoothly during Type-I bursts as the accretion rate decreased. Beyond their thermonuclear origin, the mHz QPOs in LMXBs also show intriguing interactions with the fast variability associated with the accretion flow. In particular, 4U 1608-52 provides the observational evidence for a coupling between mHz and kHz QPOs, where the kHz QPO frequency anti-correlates with the mHz QPO modulated soft X-ray flux. This suggests that luminosity changes produced by enhanced radiation from marginally stable nuclear burning increases the radiative stresses on the inner disk, temporarily pushing the disk edge outward and thus reducing the Keplerian orbital frequency that sets the kHz QPO.
\par
Although the qualitative agreement between observations and the marginally stable burning model is compelling, several aspects of the phenomenon still lack a comprehensive quantitative explanation. Current theoretical models must still address challenges such as the localized accretion required to trigger the instability, the slow frequency drifts observed in several sources including 4U 1636-53, 4U 1608-52, and IGR J17480-2446 and the complex harmonic structures seen in IGR J17480–2446. Therefore, while the observed properties of NS-LMXBs strongly support the nuclear-burning interpretation, further advances in hydrodynamic simulations are essential to fully understanding the physical conditions and intricate variability patterns revealed by observations.
\par
In contrast to the relatively coherent phenomenology of mHz QPOs in NS-LMXBs, the behavior of these oscillations in NS-HMXBs is more complex, reflecting the higher companion mass and variable stellar wind or transient disc accretion in HMXBs as illustrated by their different locations in the Corbet diagram (Figure \ref{fig:fig18}). While mHz QPOs have been reported in numerous systems, including Cen X-3, Her X-1, IGR J19294+1816, 4U 0115+63, and V0332+53, their observational properties display remarkable diversity, defying a unified explanation. In HMXBs, the QPOs span a broader frequency range, exhibit varied energy dependencies with the rms amplitude increasing or decreasing with energy depending on the source, and show different correlations with luminosity and orbital phase. This rich phenomenology stands in stark contrast to the more uniform behavior observed in LMXBs. Despite the large number of known HMXB QPO sources, the observational data for individual systems often remain sparse, typically limited to a few detections during outbursts. This scarcity of high-quality, multi-epoch data greatly hampers efforts to understand the evolution of QPO properties and to rigorously test theoretical models against a comprehensive observational phenomenon. Consequently, the theoretical interpretation of HMXB QPOs remains dispersed, with multiple models, including the KFM, BFM, the magnetic disk precession model, and the orbiting blob model, invoked to explain different systems or even distinct QPOs within the same source. While each model can account for certain aspects of the observed phenomena, none has been proven universally successful, all of them face challenges in explaining the full range of observational features, such as the lack of expected luminosity-frequency correlations in V0332+53, the complex energy-dependent sidebands in IGR J19294+1816 and the simultaneous detection of QPOs in both the optical and X-ray bands in Her X-1. This persistent inability of standard models to provide a comprehensive framework highlights the need for a new generation of targeted observations.
\par
The mHz QPOs have also been detected in X-ray pulsars in ultraluminous X-ray sources (ULXs). For instance, in the pulsating ULX M82\,X-2, the observed QPO frequency is around $\sim 3$\,mHz; M51 ULX-7 with a QPO at $\nu$ $\approx$ 0.5 mHz; and a QPO at $\nu\approx$ 10 mHz in NGC 7793 P13 \citep{feng2010discovery}. Several theoretical models have been proposed to explain the origin of such low-frequency QPOs. One interpretation invokes the general relativistic frame-dragging effect \citep{kong2016possible,vasilopoulos2020m51}. Another recent important model is magnetically driven precession \citep{veresvarska2025wobbling}, in which the tilted magnetic field of the neutron star exerts a torque on the inner accretion flow, driving the entire inner disk region to precess and thereby producing quasi-periodic modulation in the emitted radiation. This contrasts sharply with possible origins of QPOs in LMXBs, where mechanisms such as nuclear burning may prevail. Such differences indicate that the timing behavior of accreting X-ray pulsars is not governed by a universal mechanism, but is instead dictated by the system's specific physical conditions, including the accretion rate and magnetic field strength.
\par
Future observations are poised to significantly advance our understanding of mHz QPOs in neutron star systems. Current X-ray observatories, including the Hard X-ray Modulation Telescope (Insight-HXMT) and the Einstein Probe (EP), together with the forthcoming enhanced X-ray Timing and Polarimetry mission (eXTP), will provide high precision timing data for probing these oscillations. Moreover, systematic searches in multiwavelength studies, particularly to trace accretion flows and material transfer from the companion star in the optical, and probe magnetospheric interactions and jet activity in the radio may uncover corresponding signals across multiple wavelengths, offering complementary insights into their origin. Through coordinated multi-wavelength observations and continued theoretical efforts, we anticipate achieving a more comprehensive understanding of the physical mechanisms driving mHz QPOs in accreting X-ray pulsars.
\par
Systematic multi-wavelength searches—particularly in the optical to trace accretion flows and material transfer from the companion star, and in the radio to probe magnetospheric interactions and jet activity—may uncover corresponding signals and offer complementary insights into their origin. Through coordinated observations and continued theoretical efforts, we anticipate achieving a more comprehensive understanding of the physical mechanisms driving mHz QPOs in X-ray pulsars.

\funding{This work is supported by the NSFC (12133007) and National Key Research and Development Program of China (grant No. 2021YFA0718503 and 2023YFA1607901). }

\dataavailability{No new data were created or analyzed in this study. Data sharing is not applicable to this article. } 
\conflictsofinterest{The authors declare no conflicts of interest. } 

\section*{References}

\bibliography{references.bib}

\end{document}